\documentclass[12pt, a4paper]{article}
\pdfoutput=1
\usepackage{graphicx}
\usepackage{amssymb}
\usepackage{amsmath}
\usepackage{bm}
\usepackage[table,dvipsnames]{xcolor} 
\usepackage{cite}
\usepackage{slashed}
\usepackage{epstopdf}            
\usepackage{epsfig}
\usepackage{here}
\usepackage{comment}
\usepackage{booktabs} 
\usepackage{colortbl} 
\usepackage{wrapfig}
\usepackage{ascmac}
\usepackage{fancybox}  
\usepackage{soul} 

\renewcommand{\thefootnote}{\fnsymbol{footnote}}
\numberwithin{equation}{section} 
\def\beq#1\eeq{\begin{align}#1\end{align}}

\RequirePackage{xspace}
\def\Bbar    {\kern 0.18em\overline{\kern -0.18em B}{}\xspace}

\newcommand{\eg}{{\em e.g.}}
\newcommand{\ie}{{\em i.e.}}
\usepackage{adjustbox}
\usepackage{tabularx}
\newcolumntype{Y}{>{\centering\arraybackslash}X} 
\usepackage{booktabs} 
\usepackage{colortbl} 
\RequirePackage{xspace}
\def\Bbar    {\kern 0.18em\overline{\kern -0.18em B}{}\xspace}

\newcommand{\eq}[1]{Eq.~(\ref{#1})}

\usepackage{caption}
\usepackage{subcaption}
\definecolor{BlueViolet}{rgb}{0.2, 0.00, 0.7}
\definecolor{Blue}{rgb}{0.15, 0.00, 0.9}
\definecolor{light_blue}{rgb}{0.15, 0.35, 0.95}
\definecolor{kit_green}{rgb}{0, 
0.58823 
, 0.50980 
}
\usepackage[
colorlinks=true, linkcolor=light_blue,citecolor=light_blue,urlcolor=kit_green]{hyperref} 
\usepackage[hyphenbreaks]{breakurl} 

\begin{document}
\sloppy 

\begin{titlepage}

\begin{center}

\hfill{P3H--23--025, TTP23--013}

\vskip .1in

{\Large{\bf Current Status of the Muon $\boldsymbol{g-2}$ Interpretations\\[0.3 cm]
within Two-Higgs-Doublet Models}}\\

\vskip .3in

\makeatletter\g@addto@macro\bfseries{\boldmath}\makeatother

{
{\bf 
Syuhei Iguro,$^{\rm (a,b)}$ 
Teppei Kitahara,$^{\rm (c,d)}$
Martin S. Lang,$^{\rm (a)}$\\[0.25 cm]
and Michihisa Takeuchi$^{\rm (e,f)}$
}
}
\vskip .3in

\begingroup\small
\begin{tabbing}
$^{\rm (a)}$ \= {\it 
Institute for Theoretical Particle Physics (TTP), Karlsruhe Institute of Technology}\\
\> {\it  (KIT), Wolfgang-Gaede-Str.\,1, 76131 Karlsruhe, Germany}
\\[0.2em]
$^{\rm (b)}$ \> {\it Institute for Astroparticle Physics (IAP),
KIT, 
Hermann-von-Helmholtz-Platz 1,}\\
\> {\it 76344 Eggenstein-Leopoldshafen, Germany}
\\[0.2em]
$^{\rm (c)}$ \> {\it 
Kobayashi-Maskawa Institute for the Origin of Particles and the Universe,}\\
\> {\it  Nagoya University, Nagoya 464--8602, Japan}
\\[0.2em]
$^{\rm (d)}$ \> {\it 
CAS Key Laboratory of Theoretical Physics, Institute of Theoretical Physics, } \\
\> {\it Chinese Academy of Sciences, Beijing 100190, China}
\\[0.2em]
$^{\rm (e)}$ \> {\it 
School of Physics and Astronomy, Sun Yat-sen University, 519082 Zhuhai, China}
\\[0.2em]
$^{\rm (f)}$ \> {\it 
Graduate School of Information Science and Technology,
Osaka University,}\\
\> {\it Suita, Osaka 565-0871, Japan}
\end{tabbing}
{\href{mailto:igurosyuhei@gmail.com}{igurosyuhei@gmail.com}, \href{mailto:teppeik@kmi.nagoya-u.ac.jp}{teppeik@kmi.nagoya-u.ac.jp},
\href{mailto:martin.lang2@kit.edu}{m.lang@kit.edu},
\href{mailto:takeuchi@mail.sysu.edu.cn}{takeuchi@mail.sysu.edu.cn}}\\
\endgroup 


\end{center}

\begin{abstract}
In this article, we review and update implications of the muon anomalous magnetic moment (muon $g-2$) anomaly for two-Higgs-doublet models (2HDMs), which are classified according to imposed symmetries and their resulting Yukawa sector.
In the minimal setup, the muon $g-2$ anomaly can be accommodated by the type-X (lepto-philic) 2HDM, flavor-aligned 2HDM (FA2HDM), muon-specific 2HDM ($\mu$2HDM), and $\mu\tau$-flavor violating 2HDM.
We summarize all relevant experimental constraints from  high-energy collider experiments and flavor experiments, as well as the theoretical constraints from 
the perturbative unitarity and vacuum stability bounds, to these 2HDMs in light of the muon $g-2$ anomaly.
We clarify the available parameter spaces of these 2HDMs and investigate how to probe the remaining parameter regions in future experiments.
In particular, we find that, due to the updated $B_s\to\mu^+ \mu^-$ measurement, the remaining parameter region of the FA2HDM is almost equivalent to the one of the type-X 2HDM.
Furthermore, based on collider simulations, we find that the type-X 2HDM is excluded and the $\mu$2HDM scenario will be covered with the upcoming Run\,3 data.
\end{abstract}
{\sc Keywords: Muon $g-2$, 2HDM, Large Hadron Collider, Flavor Physics} 
\end{titlepage}

\setcounter{page}{1}
\renewcommand{\thefootnote}{\#\arabic{footnote}}
\setcounter{footnote}{0}

\hrule
\tableofcontents
\vskip .2in
\hrule
\vskip .4in


\section{Introduction}
\label{sec:intro}

The muon anomalous magnetic moment (the muon $g-2$), defined by $a_{\mu} =  (g_{\mu}-2)/2$, is known as a long-standing anomaly in the Standard Model (SM).
Based on the white paper recommended values \cite{Aoyama:2020ynm,Aoyama:2012wk,Aoyama:2019ryr,Czarnecki:2002nt,Gnendiger:2013pva,Davier:2017zfy,Keshavarzi:2018mgv,Colangelo:2018mtw,Hoferichter:2019mqg,Davier:2019can,Keshavarzi:2019abf,Kurz:2014wya,Melnikov:2003xd,Masjuan:2017tvw,Colangelo:2017fiz,Hoferichter:2018kwz,Gerardin:2019vio,Bijnens:2019ghy,Colangelo:2019uex,Blum:2019ugy,Colangelo:2014qya}, the SM prediction deviates from the experimental values measured at Brookhaven \cite{Muong-2:2002wip,Muong-2:2004fok,Muong-2:2006rrc} and at Fermilab \cite{Muong-2:2021ojo}, 
\begin{align}
\Delta a_\mu=a_\mu^{\rm{exp}}-a_\mu^{\rm{SM}} = \left( 25.1\pm5.9 \right)\times 10^{-10}\,,
\label{eq:anomaly}
\end{align}
at a significance of 4.2$\,\sigma$.
The measured value will be more accurately (and independently) checked by further runs of the Fermilab experiment and by the upcoming J-PARC experiment, which is based on a different measurement technique \cite{Mibe:2011zz, Abe:2019thb}.

One should note that the recent evaluation of the leading-order hadronic vacuum polarization (HVP) contribution to the muon $g-2$, based on a lattice QCD$+$QED simulation \cite{Borsanyi:2020mff}, casts doubt on the SM prediction of the white paper in \eq{eq:anomaly}, giving $\Delta a_\mu = (10.7 \pm 7.0)\times 10^{-10}$, which is consistent with the measured data $a_\mu^{\rm exp}$.
This lattice evaluation has been partially confirmed by different lattice collaborations 
\cite{Ce:2022kxy,Alexandrou:2022amy,Bazavov:2023has,Blum:2023qou}, using window observables for the HVP \cite{RBC:2018dos,Colangelo:2022vok}.
The lattice average based on these window observables provides $\Delta a_\mu = (18.3 \pm 5.9)\times 10^{-10}$, corresponding to a significance of $3.1\,\sigma$ \cite{latticeave2023}.
However, these lattice results are incompatible with $e^+e^-\to\pi^+ \pi^-$ data \cite{Crivellin:2020zul,Keshavarzi:2020bfy,Colangelo:2020lcg} at the $3.8\,\sigma$ level \cite{latticeave2023}.\footnote{Very recently, the CMD-3 collaboration reported new data for the $e^+e^-\to\pi^+ \pi^-$ cross section \cite{CMD-3:2023alj}.
The result is consistent with the lattice simulation \cite{Borsanyi:2020mff}.
However, the cross section is incompatible with other precision data by the KLOE \cite{KLOE-2:2017fda} and BaBar \cite{BaBar:2012bdw} collaborations, as well as the previous data of the prior CMD-2 collaboration \cite{CMD-2:2006gxt} without sufficient explanation.}

In this article, we assume that 
the muon $g-2$ anomaly in \eq{eq:anomaly} is a hint of new physics (NP) beyond the SM and suppose $a_\mu^{\rm NP} = ( 25.1\pm5.9 )\times 10^{-10}$.
It is known that many kinds of NP models can accommodate the muon $g-2$ anomaly, see the recent review paper \cite{Athron:2021iuf}.\footnote{See Ref.\,\cite{Lindner:2016bgg} for a previous review.} 
We will discuss NP interpretations in the context of two-Higgs-doublet models (2HDMs) in this review article.

In the last decade, our understanding of the Higgs sector has significantly been renewed after the discovery of a 125\,GeV scalar boson, with null results in searches for additional Higgs bosons at the large hadron collider (LHC).
In the meantime, precision measurements have been improved greatly.
For instance, the Yukawa couplings to third-generation fermions have been determined at the $10\%$ level, and furthermore the muon and charm Yukawa couplings were started to be measured.
With the larger statistics in the high-luminosity phase of the LHC (HL-LHC), the LHC will gain access to the triple-Higgs coupling and thus will determine the shape of the Higgs potential.

However, the fundamental question of whether the discovered 125\,GeV Higgs is the only scalar boson or just one of several scalars remains unanswered.
For example, one of the minimal extensions of the SM features an additional Higgs doublet, the so-called 2HDM \cite{Glashow:1976nt}, which naturally accommodates electroweak (EW) precision tests and has rich phenomenology.
The 2HDM appears as a low-energy effective scalar sector of many UV-complete theories, \eg, the minimal supersymmetric standard model (MSSM) \cite{Martin:1997ns,Degrassi:2002fi}, left-right model \cite{Mohapatra:1974gc}, Pati-Salam model \cite{Pati:1974yy}, little Higgs model \cite{Arkani-Hamed:2002ikv,Low:2002ws}, and so on.

It is known that the muon $g-2$ anomaly in \eq{eq:anomaly} is the same size as the EW contributions, which implies that an additional $\mathcal{O}$(100)\,GeV electroweakly-interacting particle could possibly explain the discrepancy.
The number of variants of the 2HDM that can accommodate the muon $g-2$ anomaly whilst evading existing experimental constraints is limited, and hence we can take a bottom-up approach in this review article.

Within the 2HDMs, the suppression of flavor changing neutral currents (FCNCs) is achieved once a discrete symmetry is imposed under which the two Higgs doublets and fermions carry suitable charges \cite{Glashow:1976nt}.\footnote{The suppression of the FCNCs can also be realized by using an additional $U(1)$ gauge symmetry instead of a discrete symmetry, see \eg, Refs.~\cite{Ko:2012hd,Banik:2023ecr}.}
A famous example is a discrete $\mathbb{Z}_2$ symmetry.
With the ad-hoc $\mathbb{Z}_2$ symmetry, depending on the assignment of the $\mathbb{Z}_2$ charges to the SM fermions, four different types of Yukawa interactions are allowed \cite{Barger:1989fj,Grossman:1994jb}, known as type-I, type-II, type-X and type-Y 2HDMs.
Among the four scenarios, only the type-X 2HDM can explain the $g-2$ anomaly without conflicting with experimental constraints \cite{Broggio:2014mna,Wang:2014sda}.
In this model, the two-loop Barr-Zee diagram, in which a light CP-odd Higgs and the tau lepton are in the loop, gives the dominant contribution to the muon $g-2$~\cite{Cao:2009as,Abe:2015oca,Hektor:2015zba,Crivellin:2015hha,Chun:2015hsa,Han:2015yys,Chun:2016hzs,Cherchiglia:2017uwv,Wang:2018hnw,Chun:2019oix,Sabatta:2019nfg,Jueid:2021avn,Ferreira:2021gke,Atkinson:2022qnl}.

An alternative method to eliminate the FCNCs is assuming 
flavor alignment of the Yukawa matrices for each type of right-handed fermions.
Such a model is called the flavor-aligned 2HDM (FA2HDM).
In the FA2HDM, all Yukawa interactions are proportional to the corresponding fermion mass matrix \cite{Pich:2009sp}.
Since the model contains the type-X 2HDM as a certain limit, it is also an interesting candidate for explaining the muon $g-2$ anomaly.
In addition to the tau-lepton Barr-Zee diagram,
a top-quark Barr-Zee diagram can also contribute to the muon $g-2$ in the FA2HDM \cite{Ilisie:2015tra,Han:2015yys}.

More complicated discrete symmetries, $\mathbb{Z}_4$, are also discussed.
In the muon-specific 2HDM ($\mu$2HDM) based on a $\mathbb{Z}_4$ symmetry, the additional Higgs doublet strongly couples only to the muon, without the FCNCs \cite{Abe:2017jqo}.
Another viable method introduces a lepton-flavor-violating scalar particle.
The magnetic dipole operator requires a chirality flip, which corresponds to the muon mass within the flavor-conserving scenarios.
On the other hand, if a model contains a $\mu\tau$-flavor violating vertex with a neutral particle, the chirality of the virtual tau lepton can be flipped instead and the one-loop contribution is enhanced by a factor of $m_\tau/m_\mu\simeq17$ compared to the flavor-conserving scenarios~\cite{Nie:1998dg,Diaz:2000rsq,Baek:2001kca,Iltan:2001nk,Wu:2001vq,Assamagan:2002kf,Omura:2015nja,Omura:2015xcg,Dev:2017ftk,Iguro:2018qzf,Abe:2019bkf,Wang:2019ngf,Iguro:2019sly,Crivellin:2019dun,Iguro:2020rby,Jana:2020pxx,Wang:2021fkn,Ghosh:2021jeg,Hou:2021sfl,Han:2022juu,Asai:2022uix}.
A $\mathbb{Z}_4$ symmetry can naturally realize such a model \cite{Abe:2019bkf}, and we call it the $\mu\tau$-flavor violating 2HDM ($\mu\tau$2HDM).
In this review article, in light of the muon $g-2$ anomaly, we update the status of these four possibilities: type-X 2HDM, FA2HDM, $\mu$2HDM, and $\mu\tau$2HDM.
Variants of the 2HDM with even larger discrete symmetries can also be conceived; however, their implications for the muon $g-2$ anomaly are mostly the same as in the $\mu$2HDM and $\mu\tau$2HDM.

In this paper we present the following updates and findings:
\begin{itemize}
 \item {Reinterpreting the latest charginos and neutralinos search based on the Run\,2 full data, we find that the tau-rich signature at the LHC excludes the muon $g-2$ explanation within the type-X 2HDM. }
 \item {Including the recent CMS $B_s\to \mu^+ \mu^-$ measurement, which is consistent with the SM prediction, we find that the parameter space of the FA2HDM is similar to the one in the type-X 2HDM in terms of the muon $g-2$. 
 Therefore the FA2HDM explanation of the anomaly is also excluded. }
 \item {We update the collider constraint on $\mu$2HDM by using results of the latest multi-lepton search.
 By incorporating also the theoretical constraint from the Landau pole, we find that the current central value of the muon $g-2$ anomaly indicates that the Landau pole scale is less than $\simeq 5\,$TeV in the model. 
 Moreover, we obtain a projection for future sensitivity and find that  500\,fb$^{-1}$ of data will cover the 1\,$\sigma$ region of the muon $g-2$ anomaly once we impose the scale to be larger than $\simeq 5\,$TeV.
 We also point out that a muon-flavor exclusive multi-lepton search can improve the sensitivity to the model.}
\end{itemize}

The outline of the paper is as follows.
In Sec.~\ref{sec:model},
all kinds of minimal 2HDMs which can resolve the muon $g-2$ anomaly are introduced.
In Sec.~\ref{sec:pheno}, we examine various experimental and theoretical constraints on their parameter spaces and discuss the future experimental prospects for them.
Finally, conclusions are drawn in Sec.~\ref{sec:summary}.
In the Appendix, we collect all relevant formulae for the analyses of this article.

\section{Two-Higgs-doublet models}
\label{sec:model}

First, we introduce two different bases in the various 2HDMs: the so-called $\mathbb{Z}_2$ basis and the Higgs basis, which are mathematically equivalent.\footnote{%
For the parameter conversions between the two bases, see Appendices of Refs.~\cite{Aiko:2020atr,Iguro:2022fel} for the Higgs potential and Ref.~\cite{Iguro:2017ysu} for the Yukawa sector.}
The $\mathbb{Z}_2$ basis respects charge assignments of the discrete symmetry.
One can straightforwardly track the free parameters of the model in the $\mathbb{Z}_2$ basis.
On the other hand, the Higgs basis which will be used in this article can parametrize more general 2HDMs.
In particular, if one assumes the alignment of the SM Higgs boson, any calculations become significantly simpler in the Higgs basis.

Within the 2HDMs, when the Higgs potential is minimized at the EW symmetry breaking vacuum, both neutral components of Higgs doublets acquire vacuum expectation values (VEVs) in general.
In the $\mathbb{Z}_2$ basis, the Higgs potential is given by 
\begin{align}
  V(\Phi_1, \Phi_2)&=m_{11}^2 \Phi_1^\dagger \Phi_1 +m_{22}^2 \Phi_2^\dagger \Phi_2-\left(m_{12}^2\Phi_1^\dagger \Phi_2+{\rm h.c.}
  \right)\nonumber \\
&\quad +\frac{\lambda_1^{Z_2}}{2}(\Phi_1^\dagger \Phi_1)^2+\frac{\lambda_2^{Z_2}}{2}(\Phi_2^\dagger \Phi_2)^2+\lambda_3^{Z_2}(\Phi_1^\dagger \Phi_1)(\Phi_2^\dagger \Phi_2)
+\lambda_4^{Z_2} (\Phi_1^\dagger \Phi_2)(\Phi_2^\dagger \Phi_1)\nonumber \\
&\quad +\frac{1}{2}\left(
\lambda_5^{Z_2}(\Phi_1^\dagger \Phi_2)^2+{\rm h.c.}\right)\,,
\end{align}
where in general $m^2_{11},\, m^2_{22}$ and $\lambda_{1-4}^{Z_2}$ are real parameters, while $m_{12}^2$ and $\lambda_5^{Z_2}$ are complex ones.
In this article, we assume absence of CP violation in the Higgs potential for simplicity, so that $m_{12}^2$ and $\lambda_5^{Z_2}$ are treated as real parameters as well.
The two Higgs doublets in the $\mathbb{Z}_2$ basis are defined as 
\begin{eqnarray}
  \Phi_1 =\left(
  \begin{array}{c}
    \omega_1^+\\
    \frac{v_1+h_1+ \mathrm{i} z_1}{\sqrt{2}}
  \end{array}
  \right)\,,\quad 
  \Phi_2=\left(
  \begin{array}{c}
    \omega_2^+\\
    \frac{v_2+h_2+\mathrm{i} z_2}{\sqrt{2}}
  \end{array}
  \right)\,.
\end{eqnarray}
The VEVs $v_1,\,v_2$ can be taken to be real and positive and need to satisfy $v=\sqrt{v_1^2+v_2^2}\simeq246\,$GeV in order to reproduce the masses of the weak gauge bosons.
The ratio of the VEVs is represented by $\tan\beta = v_2/v_1~(0\leq \beta \leq \pi/2)$.

By taking a certain linear combination of $\Phi_{1,2}$, one can always work in the Higgs basis \cite{Georgi:1978ri,Donoghue:1978cj,Davidson:2005cw} where only one Higgs doublet obtains a VEV as 
\begin{eqnarray}
  \left(
  \begin{array}{c}
    H_1\\
    H_2
  \end{array}
  \right) =\left(
  \begin{array}{cc}
    \cos \beta &\sin\beta\\
    -\sin \beta &\cos\beta
\end{array}
  \right)
    \left(
  \begin{array}{c}
    \Phi_1\\
    \Phi_2
  \end{array}
  \right)\,.
\label{BasisInterplay}
\end{eqnarray}
In the Higgs basis, the doublets can be decomposed as
\begin{eqnarray}
  H_1 =\left(
  \begin{array}{c}
    G^+\\
    \frac{v+h+ \mathrm{i} G}{\sqrt{2}}
  \end{array}
  \right),~~~
  H_2=\left(
  \begin{array}{c}
    H^+\\
    \frac{H+ \mathrm{i} A}{\sqrt{2}}
  \end{array}
  \right)\,,
\label{HiggsBasis}
\end{eqnarray}
where the fields $G^+$, $H^+$, $h$, $H$, $G$, and $A$ are linear combinations of $\omega_{1,2}^+, h_{1,2}$, and $z_{1,2}$.
The Nambu-Goldstone bosons of the spontaneously broken EW gauge symmetry are denoted by $G^\pm$ and $G$, and $H^\pm$ denotes an additional charged Higgs boson, while $A$ is a neutral CP-odd Higgs boson.
In principle, the CP-even scalars $h$ and $H$ in the doublets mix with an angle $\alpha$ to constitute the mass eigenstates.
However, since the LHC found that the interactions of the observed scalar boson are currently consistent with the SM Higgs expectations, we consider the case where the mixing between CP-even scalars is negligible corresponding to a conservative choice, \ie, $\sin\left(\beta - \alpha\right) \simeq 1$, such that $h$ and $H$ are promoted to mass eigenstates.\footnote{If non-zero mixing is considered, we have more stringent constraints on the model from, \eg, $h\to\tau^+\tau^-$.}
Since experimental constraints are commonly weakened in the Higgs alignment limit, we expect the study in this article to be conservative.

In the Higgs basis, the scalar potential is given by
\begin{align}
  V(H_1, H_2)&=M_{11}^2 H_1^\dagger H_1+M_{22}^2 H_2^\dagger H_2-\left(M_{12}^2 H_1^\dagger H_2+{\rm h.c.}
  \right)\nonumber \\
&\quad +\frac{\lambda_1}{2}(H_1^\dagger H_1)^2+\frac{\lambda_2}{2}(H_2^\dagger H_2)^2+\lambda_3(H_1^\dagger H_1)(H_2^\dagger H_2)
+\lambda_4 (H_1^\dagger H_2)(H_2^\dagger H_1)\nonumber \\
&\quad +\left\{
\frac{\lambda_5}{2}(H_1^\dagger H_2)^2+\left[
\lambda_6 (H_1^\dagger H_1)+\lambda_7 (H_2^\dagger H_2)\right] (H_1^\dagger H_2)+{\rm h.c.}\right\}\,.
\label{eq:Higgsbasis}
\end{align}
By matching to the Higgs potential in the $\mathbb{Z}_2$ basis, one obtains
\begin{align}
    \lambda_{6} 
&= -\frac{1}{2} \sin{2\beta}\left(\cos^{2}{\beta}\lambda^{Z_2}_{1} -  \sin^{2}{\beta} \lambda^{Z_2}_{2} -  \cos{2\beta}\lambda^{Z_2}_{345} \right)\,, 
\label{eq:l6}\\
\lambda_{7} 
&= -\frac{1}{2} \sin {2\beta}\left( \sin^{2} {\beta} \lambda_{1}^{Z_2}- \cos^{2}{\beta}\lambda_{2}^{Z_2} + \cos{2\beta}\lambda_{345}^{Z_2}\right)\,,
\label{eq:l7}
\end{align}
with $\lambda_{345}^{Z_2} = \lambda_{3}^{Z_2}+\lambda_{4}^{Z_2}+\lambda_{5}^{Z_2}$.
We note that the Higgs alignment condition corresponds to $\lambda_6=0$ at the renormalization scale $\mu\approx m_W$ in the Higgs basis, which leads to $M_{12}^2=0$ under the stationary condition.

The scalar mass spectrum is important for our discussion and hence it is useful to show the mass relations in the Higgs alignment limit in the Higgs basis:
\begin{align}
  \begin{aligned}
  m_h^2& = \lambda_1 v^2\,, \quad 
  m_A^2 = M_{22}^2+\frac{\lambda_3+\lambda_4-\lambda_5}{2} v^2\,, \\
  m_H^2& =m_A^2+\lambda_5 v^2\,, \quad
  m_{H^\pm}^2 = m_A^2 - \frac{\lambda_4-\lambda_5}{2} v^2\,.
  \label{eq:Higgs_spectrum}
  \end{aligned}
\end{align}
 The mass differences among the neutral scalars are crucial parameters to discuss the muon $g-2$ anomaly and constraints from collider physics.
The numerical relation is given as
\begin{align}
\Delta_{H-A}& = m_H-m_A\,\simeq 60\left(\frac{\lambda_5 }{1.0}\right)\left(\frac{1000\,{\rm{GeV}}}{m_H+m_A}\right) {\rm{GeV}}\,.
\label{Eq:mass_l5}
\end{align}
We also define 
\begin{align}
\lambda_{hAA}=\lambda_{3}+\lambda_{4}-\lambda_{5}\,,
\end{align}
for later convenience, which corresponds to the $hAA$ coupling in the Higgs alignment limit.

The most general Yukawa sector of the 2HDM in the fermion mass eigenbasis is given as
\begin{eqnarray}\label{eq:yukferbasis1}
  - \mathcal{L} &=&\overline{Q}_L^i H_1 y_d^i d_R^i +\overline{Q}_L^i H_2 \rho_d^{ij} d_R^j+\overline{Q}_L^i (V^\dagger)^{ij} \widetilde{H}_1 y_u^j u_R^j
  +\overline{Q}_L^i (V^\dagger)^{ij} \widetilde{H}_2 \rho_u^{jk}u_R^k \nonumber\\
  &&+\overline{L}_L^i H_1 y^i_e e_R^i +\overline{L}_L^i H_2 \rho^{ij}_e e_R^j + \rm{h.c.} \,.
\label{yukawas}
\end{eqnarray}
Here $i,j$ denote flavor indices and $Q=(V^\dagger u_L,d_L)^T$ and $L=(U \nu_L, e_L)^T$ are $\rm{SU}(2)_L$ doublets, where $V$ and $U$ are the Cabbibo-Kobayashi-Maskawa (CKM) \cite{Cabibbo:1963yz,Kobayashi:1973fv} and the Pontecorvo-Maki-Nakagawa-Sakata (PMNS) \cite{Pontecorvo:1957qd,Maki:1962mu} matrices, respectively.
In writing Eq.~(\ref{eq:yukferbasis1}) we have assumed that neutrino masses are explained by the seesaw mechanism introducing heavy right-handed neutrinos, so that in the low-energy effective theory the left-handed neutrinos have a $3\times 3$ Majorana mass matrix.
Note that the Yukawa couplings $y_f$ are expressed in terms of the fermion masses $m_f$, $y_f=\sqrt{2}m_f/v$.
On the other hand, the Yukawa couplings $\rho_f^{ij}$ are a priori arbitrary $3\times3$ complex matrices and can in general be sources of flavor violation mediated by additional Higgs bosons at tree level.
In the Higgs alignment limit, the interactions of $H_1$ are exactly the same as the ones of the SM Higgs doublet.

Following the notation of Ref.~\cite{Iguro:2017ysu}, in terms of the mass eigenstates of the Higgs bosons, the Yukawa interactions are represented by
\begin{align}
  -\mathcal{L}&=\sum_{f=u,d,e}\sum_{\phi=h,H,A} y^f_{\phi i j}\overline{f}_{Li} \phi f_{Rj}+{\rm h.c.}
  \nonumber\\
  &\quad +\overline{\nu}_{Li} (U^\dagger \rho_e)^{ij}  H^+ e_{Rj}
  +\overline{u}_i(V\rho_d P_R-\rho_u^\dagger V P_L)^{ij} H^+d_j+{\rm h.c.}\,,
\end{align}
where
\begin{align}
  y^f_{hij}&= \frac{1}{\sqrt{2}} y_{f}^i \delta_{ij}= \frac{m_{f_i}}{v}\delta_{ij}\,,\quad 
  y^f_{Hij}=\frac{1}{\sqrt{2}}\rho_f^{ij}\,, \quad 
  y^f_{Aij}=
  \left\{
  \begin{array}{c}
    -\frac{\mathrm{i} }{\sqrt{2}}\rho_f^{ij}~~~{\rm for}~f=u\,,\\
    \frac{ \mathrm{i}}{\sqrt{2}}\rho_f^{ij}~~~{\rm for}~f=d, e\,.
  \end{array}
  \right.
  \label{yukawa}
\end{align}

The off-diagonal components of the Yukawa couplings $\rho_f^{ij}$ ($i \neq j$) induce FCNCs from decays of the scalar bosons and Higgs-mediated processes.
The absence of such FCNCs at experiments\footnote{%
The recent ATLAS and CMS results both show hints of Higgs lepton-flavor violating decays \cite{CMS:2021rsq,ATLAS:2023mvd,CMS:2023pqk}.
In order to address these slight deviations, one needs a small neutral-scalar mixing.} 
motivates us to impose a discrete symmetry which distinguishes the two Higgs doublets $\Phi_{1,2}$.
The Yukawa structure of $\rho_f^{ij}$ depends on the charge assignment of the Higgs doublets and fermions as well, which we will classify in the following sections.

\subsection{Type-X 2HDM}
\label{sec:tX_2HDM}
We start with the type-X 2HDM.
This is one of the four types of 2HDMs with softly-broken $\mathbb{Z}_2$ symmetry and thus naturally suppresses FCNCs.
The $\mathbb{Z}_2$ assignment is summarized in Table~\ref{tab:Z-charge}.
\begin{table}[t]
\begin{center}
    \begin{tabularx}{0.8 \textwidth}{cYYYYYcc} 
    \toprule
     &$\Phi_1$&$\Phi_2$&$Q$&$u_R$&$d_R$&$(L_e,L_\mu,L_\tau)$&$(e_R, \mu_R, \tau_R)$ \\ \hline 
   type-I &$0$&$1$&$0$&$1$&$1$&$0$&$1$\\ 
   type-II &$0$&$1$&$0$&$1$&$0$&$0$&$0$\\ 
   type-X &$0$&$1$&$0$&$1$&$1$&$0$&$0$\\ 
   type-Y &$0$&$1$&$0$&$1$&$0$&$0$&$1$\\ \hline
   $\mu$2HDM &$2$&$0$&$0$&$0$&$0$&$(0,-1,0)$&$(0,1,0)$ \\
   $\mu\tau$2HDM
   &$2$&$0$&$0$&$0$&$0$&$(0,1,-1)$&$(0,1,-1)$\\
   \bottomrule
  \end{tabularx}
  \caption{The charge assignments under the discrete symmetry: 
 the matter fields are transformed as $\psi \to \exp(2 \pi q_\psi {\rm{i}}/N) \, \psi$ under the discrete $\mathbb{Z}_N$ symmetry.
 We show the $\mathbb{Z}_2$ charge assignments for the type-I, II, X, and Y 2HDMs ($0$ is $\mathbb{Z}_2$-even, while $1$ is $\mathbb{Z}_2$-odd) \cite{Aoki:2009ha} and 
 the $\mathbb{Z}_4$ charge assignments for the $\mu$2HDM \cite{Abe:2017jqo} and the $\mu\tau$2HDM \cite{Abe:2019bkf}.
 }
 \label{tab:Z-charge}
\end{center}
\end{table}
Out of the four $\mathbb{Z}_2$-symmetric 2HDMs, only the type-X 2HDM can resolve the muon $g-2$ anomaly because in all three other 2HDMs a large contribution to $a_\mu^{\rm NP}$ is not allowed by flavor and collider constraints and the perturbative unitarity bound.
More explicitly, in the type-I and type-Y 2HDMs, both $\rho_u^{ii}$ and $\rho_e^{ii}$ are proportional to $\cot \beta$.
Therefore, the upper limit on $\rho_u^{tt}$ from perturbativity indirectly sets an upper limit on $\rho_e^{\mu\mu}$, suppressing any potential contributions to $a_\mu^{\rm NP}$.\footnote{See also Ref.~\cite{Atkinson:2022pcn}.}
On the other hand, in the type-II and type-X 2HDMs, $\rho_e^{ii}$ is proportional to $\tan\beta$ while $\rho_u^{ii}$ remains proportional to $\cot\beta$.
It is noted that the type-II and type-X 2HDMs cannot accommodate the muon $g-2$ anomaly at one-loop level with a light CP-even Higgs.\footnote{See also Ref.~\cite{Atkinson:2021eox}.}
If one tries to explain $\Delta a_\mu$ with $\rho_e^{\mu\mu}=\mathcal{O}(1)$, $\rho_e^{\tau\tau}$ immediately becomes non-perturbative because $\rho_e^{\tau\tau}/\rho_e^{\mu\mu}$ is fixed by $m_\tau/m_\mu$.
Instead, the contributions to $a_\mu^{\rm NP}$ can be dominated by the so-called Barr-Zee diagram \cite{Barr:1990vd} at two-loop level with a light CP-odd Higgs~\cite{Cao:2009as,Abe:2015oca,Hektor:2015zba,Crivellin:2015hha,Chun:2015hsa,Han:2015yys,Chun:2016hzs,Cherchiglia:2017uwv,Wang:2018hnw,Chun:2019oix,Sabatta:2019nfg,Jueid:2021avn,Ferreira:2021gke,Atkinson:2022qnl}.
We exhibit the Feynman diagrams in Fig.\,\ref{Fig:diag-2} and the formula in Appendix~\ref{app:muong-2}.

The difference between the type-II and type-X 2HDMs is the down-type quark Yukawa couplings to the extra Higgs bosons.
In the type-II 2HDM, both the down-type quark and charged lepton couplings are enhanced by $\tan\beta$ at the same time.
Therefore, the model is severely constrained by $B$-meson flavor physics and direct searches for extra Higgs bosons.
The rare radiative decay $b\to s\gamma$ gives a lower mass limit for the charged Higgs boson of $m_{H^\pm}\ge 800$\,GeV \cite{Misiak:2020vlo}.
For the muon $g-2$ anomaly, we need a $\mathcal{O}$(10--100)\,GeV light scalar at the same time.
Such a large mass difference is troublesome since the theory will be non-perturbative at less than 1\,TeV.

On the other hand, only the lepton Yukawa couplings are enhanced by $\tan\beta$ in the type-X 2HDM.
Therefore, the constraints from $B$-meson decays are weaker compared to the type-II ones.
The Yukawa structure in the type-X 2HDM is given by
\begin{align}
    \rho_u=\frac{\sqrt{2}m_{u_i}}{v} \xi^{-1}\,,\quad \rho_d=\frac{\sqrt{2}m_{d_i}}{v}\xi^{-1}\,, \quad \rho_e=-\frac{\sqrt{2}m_{e_i}}{v}\xi\,,
\end{align}
with $\xi=\tan\beta$ and all non-diagonal Yukawa couplings vanishing.
Although $\tan\beta$ is conventionally used, we will use the notation $\xi$ in this article in order to allow for an easy comparison with the other 2HDM scenarios.

\begin{figure}[t]
\begin{center}
\begin{subfigure}{0.4 \textwidth}
\includegraphics[width = \textwidth]{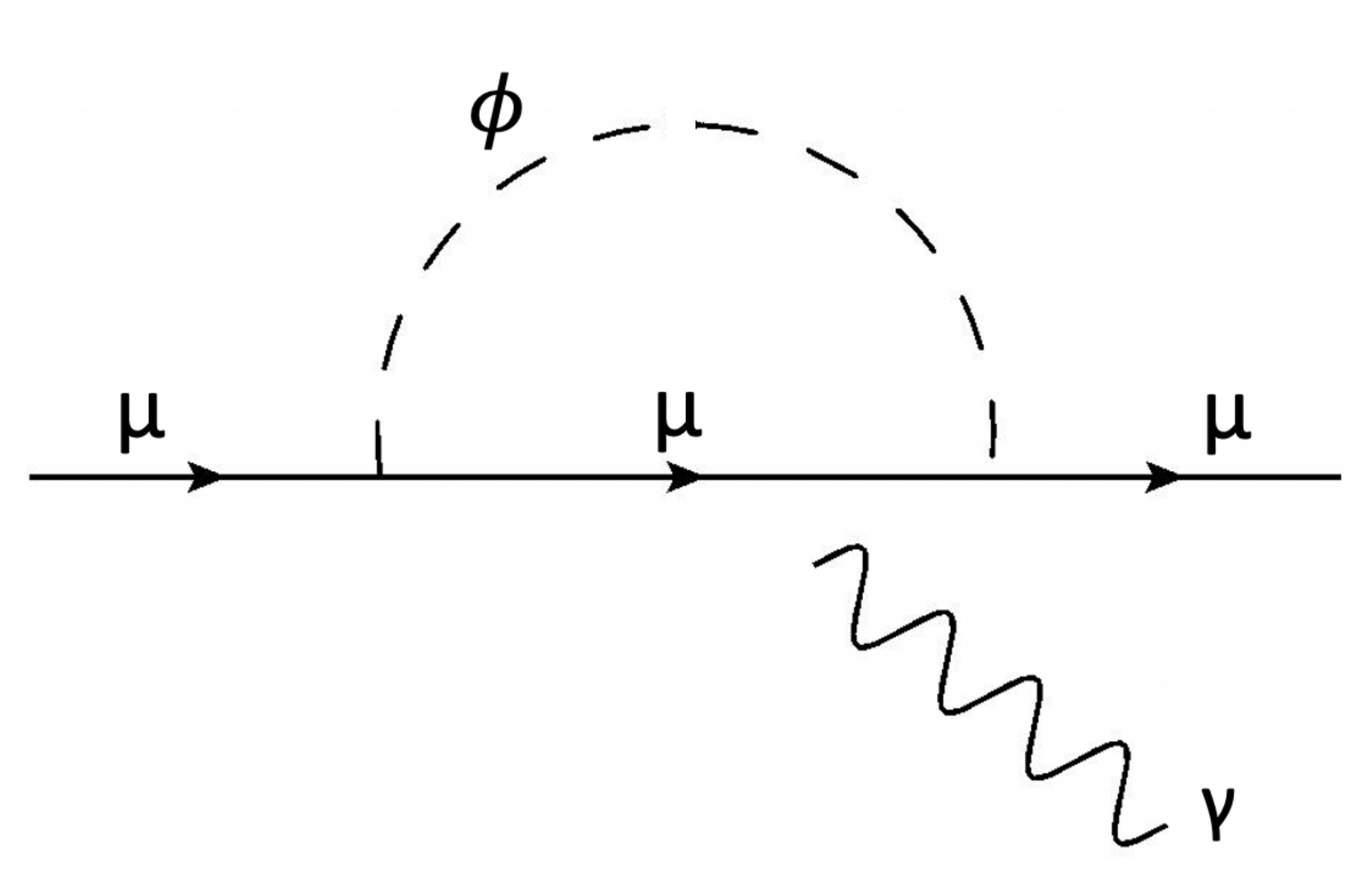}
\end{subfigure}
\qquad 
\begin{subfigure}{0.4 \textwidth}
\includegraphics[width = \textwidth]{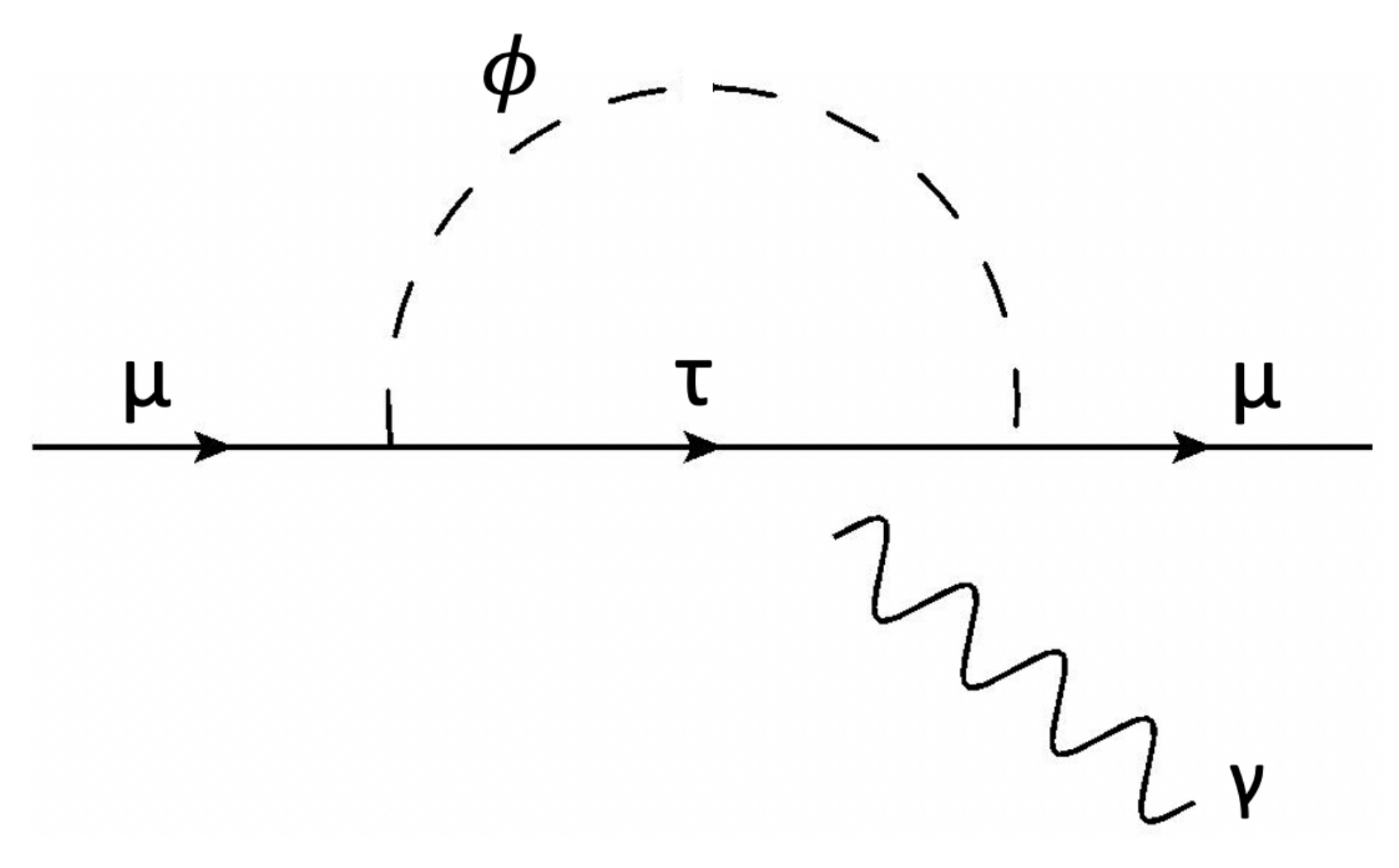}
\end{subfigure}\\[-1cm]
\begin{subfigure}{0.3 \textwidth}
\includegraphics[width = \textwidth]{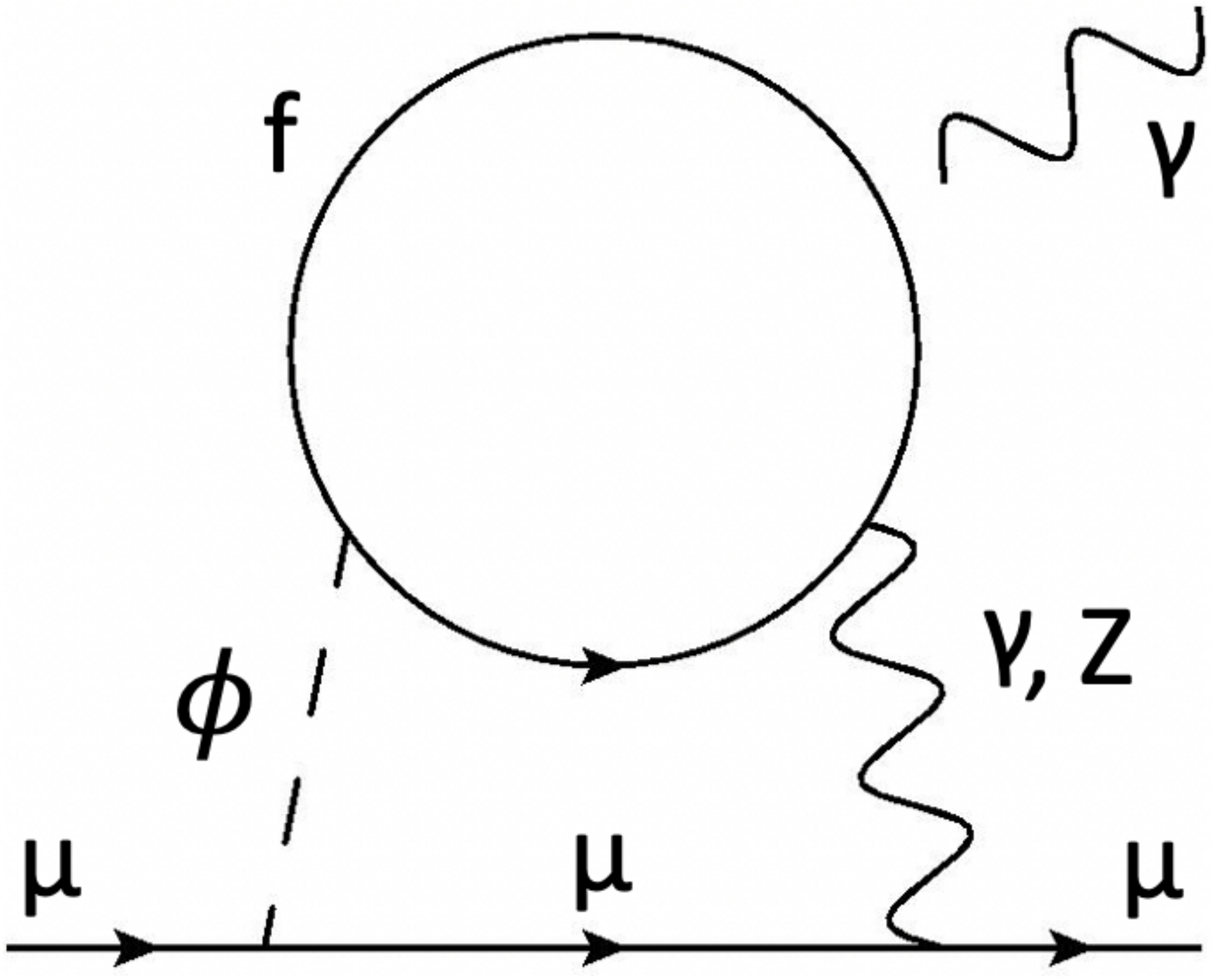}
\end{subfigure}
\vspace{-1cm}
\caption{
\label{Fig:diag-2}
The diagram of leading contribution to the muon $g-2$ in the $\mu$2HDM (left), $\mu\tau$2HDM (right), and type-X 2HDM and FA2HDM (bottom).
The bottom diagram is the two-loop Barr-Zee diagram where $f=\tau,\,t$.
} 
\end{center}
\end{figure}

\subsection{Flavor-aligned 2HDM}
\label{sec:fa_2HDM}
In the flavor-aligned 2HDM (FA2HDM), it is assumed that the Yukawa interactions of the additional scalars are proportional to mass matrices ($\rho_f \propto y_f$); both the $\rho_f$ and $ y_f$ matrices are simultaneously diagonalized in the Lagrangian of Eq.\,\eqref{yukawas} \cite{Pich:2009sp}.
Therefore, tree-level FCNCs are absent in this model.\footnote{
Even if one imposes the flavor alignment at tree-level, FCNCs are in general induced radiatively, in particularly by the renormalization group evolutions (RGEs).
However, these RGE-induced \mbox{FCNCs} are shown to be too small to be probed at current experiments \cite{Braeuninger:2010td,Gori:2017qwg}.}
Viable UV models for the flavor alignment condition are discussed in Refs.~\cite{Knapen:2015hia,Penuelas:2017ikk}.

There are three free parameters in the Yukawa interactions, $\xi_{u}$, $\xi_{d}$, and $\xi_{e}$.
In other words, the type-X scenario is a special case of the FA2HDM, in which $\xi_u = \xi_d = \xi^{-1} $ and $\xi_e = - \xi$.
The additional Yukawa couplings are given as 
\begin{align}
    \rho_u=\frac{\sqrt{2}m_{u_i}}{v} \xi_u\,,\quad \rho_d=\frac{\sqrt{2}m_{d_i}}{v}\xi_d\,, \quad \rho_e=\frac{\sqrt{2}m_{e_i}}{v}\xi_e\,.
\end{align}
In this model, the dominant contribution to $a_{\mu}^{\rm NP}$ comes from the two-loop Barr-Zee diagram with a light CP-odd Higgs and a tau lepton, while the Barr-Zee diagram with a top-quark loop can also contribute \cite{Ilisie:2015tra,Han:2015yys}.

\subsection{Muon-specific 2HDM}
\label{sec:model_m2HDM}
Another interesting scenario is the so-called muon-specific 2HDM ($\mu$2HDM).
In this model, the additional scalars dominantly couple to muons due to a $\mathbb{Z}_4$ charge assignment as summarized in Table~\ref{tab:Z-charge}.
This coupling structure helps to avoid the constraints from $\tau$ decays and loop-induced $Z$ decays \cite{Abe:2017jqo}.
The Yukawa coupling structure of this model is given as
\begin{align}
    \rho_f=\frac{\sqrt{2}m_f}{v}\xi^{-1}\,, \quad \rho_e^{\mu\mu}=-\frac{\sqrt{2}m_\mu}{v}\xi\,,
\end{align}
with $\xi=\tan\beta$, where $f$ denotes all fermions except for $\mu$.
The dominant contribution to $a_{\mu}^{\rm NP}$ stems from one-loop diagrams, see Fig.\,\ref{Fig:diag-2} (top-left), and the formula is given in Appendix~\ref{app:muong-2}.
$m_H < m_A$ or equivalently $\lambda_5 < 0$ can yield $\Delta a_\mu > 0$.
The phenomenologically interesting parameter region is $\xi \gg 1$.

\subsection{\texorpdfstring{$\mu\tau$}{mutau}-flavor violating 2HDM}
\label{sec:mutau_2HDM}
In the $\mu\tau$-flavor violating 2HDM ($\mu\tau$2HDM), a discrete $\mathbb{Z}_4$ symmetry with charge assignments shown in Table~\ref{tab:Z-charge} is imposed.
Due to the unique charge assignment, after the diagonalization of the charged lepton mass, the only flavor-violating interactions of the additional neutral scalars in the Higgs alignment limit are the $\overline{\mu}_{L/R}\tau_{R/L} H(A)$ interactions.
The Yukawa coupling structure of this model is given as
\begin{align}
    \rho_f^{\rm diag} = \frac{\sqrt{2}m_f}{v}\xi^{-1}\,,\quad \rho_e^{\mu\tau}\neq0\,,\quad \rho_e^{\tau\mu}\neq0\,,
\end{align}
with $\xi\approx\tan \beta$.
Here, $f$ denotes all fermions.
The dominant contribution to $a_{\mu}^{\rm NP}$ comes from the one-loop diagram with a virtual tau lepton in Fig.\,\ref{Fig:diag-2}, and the formula is given in Appendix~\ref{app:muong-2}.

However, even in the case of the Higgs alignment limit, this model also predicts $\tau \to \mu$ lepton-flavor violating transitions at one-loop level.
To avoid their experimental bounds, the limit $\tan\beta \to \infty$ is a natural solution, corresponding to the original discrete $\mathbb{Z}_4$ symmetry (a variant of the inert doublet model) proposed in Ref.~\cite{Abe:2019bkf}.
Since $a_{\mu}^{\rm NP}$ is insensitive to $\tan\beta$ in this model, we consider the following Yukawa coupling structure
\begin{align}
    \rho_f^{\rm diag}  = 0\,,\quad \rho_e^{\mu\tau}\neq0\,,\quad \rho_e^{\tau\mu}\neq0\,.
    \label{eq:mutaurho}
\end{align}
Then, we can safely focus on the phenomenology of the additional scalars.
Note that for realistic neutrino masses and mixing, breaking of the $\mathbb{Z}_4$ symmetry is necessary \cite{Asai:2022uix}.

\section{Current status and prospects for muon \texorpdfstring{$g-2$}{g-2} interpretations}
\label{sec:pheno}
In this main section, we discuss explanations of the muon $g-2$ anomaly based on the models introduced in Sec.~\ref{sec:model}, along with relevant flavor and collider constraints.
The future prospects at the (HL-)LHC are also discussed.
Table~\ref{tab:ModelNRC} summarizes the interesting mass range and relevant processes.

\begin{table}[t]
\begin{center}
\scalebox{0.76}{
  \begin{tabular}{c|c|c|c|c|c} 
  \toprule
    & $\Delta a_\mu $& mass range& precision & LHC & life time  \\ \hline 
   Type-X 2HDM & 2 loop & $m_A=\mathcal{O}(10)\,{\rm{GeV}}\ll m_H=m_{H^\pm}$& $h\to AA$, $Z, \tau$ decays& multi-$\tau$  & Run~2 \\ 
   FA2HDM & 2 loop & $m_A=\mathcal{O}(10)\,{\rm{GeV}}\ll m_H=m_{H^\pm}$ & $B_s\to\mu^+\mu^-$, $h\to AA$  & multi-$\tau$ & Run~2 \\ 
   $\mu$2HDM & 1 loop &900\,GeV\,$\le m_{A,\,H}\le 1000\,$GeV & $Z$ decay & multi-$\mu$  &
   Run~3
 \\ 
   $\mu\tau$2HDM& 1 loop & 500\,GeV\,$\le m_{A,\,H}\le 1600\,$GeV & $\tau\to\mu\nu\overline\nu$& $\mu^\pm\mu^\pm\tau^\mp\tau^\mp$ & HE-LHC\\ \bottomrule  
    \end{tabular}
  }
  \caption{Summary table for all 2HDM scenarios which can accommodate the muon $g-2$ anomaly.
  The second column shows the loop order of the dominant contribution to $a_{\mu}^{\rm NP}$.
  The third column lists the available mass range of scalars.
  The fourth and fifth ones show the relevant constraints from precision measurements and important processes at LHC, respectively.
  The last column summarizes how much data is needed to fully explore the parameter space where the muon $g-2$ anomaly can be solved at the 1\,$\sigma$ level.
 }
  \label{tab:ModelNRC}
\end{center}   
\end{table}

\subsection{Type-X 2HDM}
\label{sec:tX_2HDM_pheno}

The type-X model explains the muon $g-2$ with a two-loop Barr-Zee diagram in which a CP-odd scalar $A$ of $\mathcal{O}$(20--40)\,GeV propagates.
In the large $\xi$ limit, the two-loop Barr-Zee correction with a light $A$ and tau internal loop can generate a large positive contribution to $a_\mu^{\rm NP}$ since it is enhanced by $m_\tau^2/m_\mu^2$.
The Barr-Zee contribution with $H$ in the loop gives a negative contribution and thus a heavier $H$ is preferred to enhance $ a_\mu^{\rm NP}$.

In this mass range, the SM Higgs boson can decay into a pair of light CP-odd scalars, which modifies the Higgs total width.
This additional decay channel opens for $m_h\ge2m_A$ and the tree-level $h\to AA$ partial decay width is given as
\begin{align}
\Gamma(h\to AA)=\frac{\lambda_{hAA}^2 v^2}{32\pi m_h}\sqrt{1-\frac{4m_A^2}{m_h^2}}\,.
\end{align}
Recent Higgs width measurements restrict the trilinear Higgs coupling to \cite{CMS:2022ley,ATLAS:2023dnm}
\begin{align}
|\lambda_{hAA}|\le0.03\,.
\label{eq:hAA}
\end{align}
A more stringent limit $|\lambda_{hAA}|\lesssim 0.01$ is obtained for $m_A\le 21\,$GeV based on searches for $h\to AA\to\mu^+\mu^-\tau^+\tau^-$ decays \cite{CMS:2020ffa}.
Since the non-discovery of $H^\pm$ prefers large mass differences between the CP-odd and charged scalars, $\mathcal{O}(1)$ couplings in the Higgs potential are necessary, see \eq{eq:Higgs_spectrum}.
Therefore, Eq.~(\ref{eq:hAA}) requires parameter tuning at the 1$\%$ level.

\begin{figure}[t]
\begin{center}
\includegraphics[width=17em]{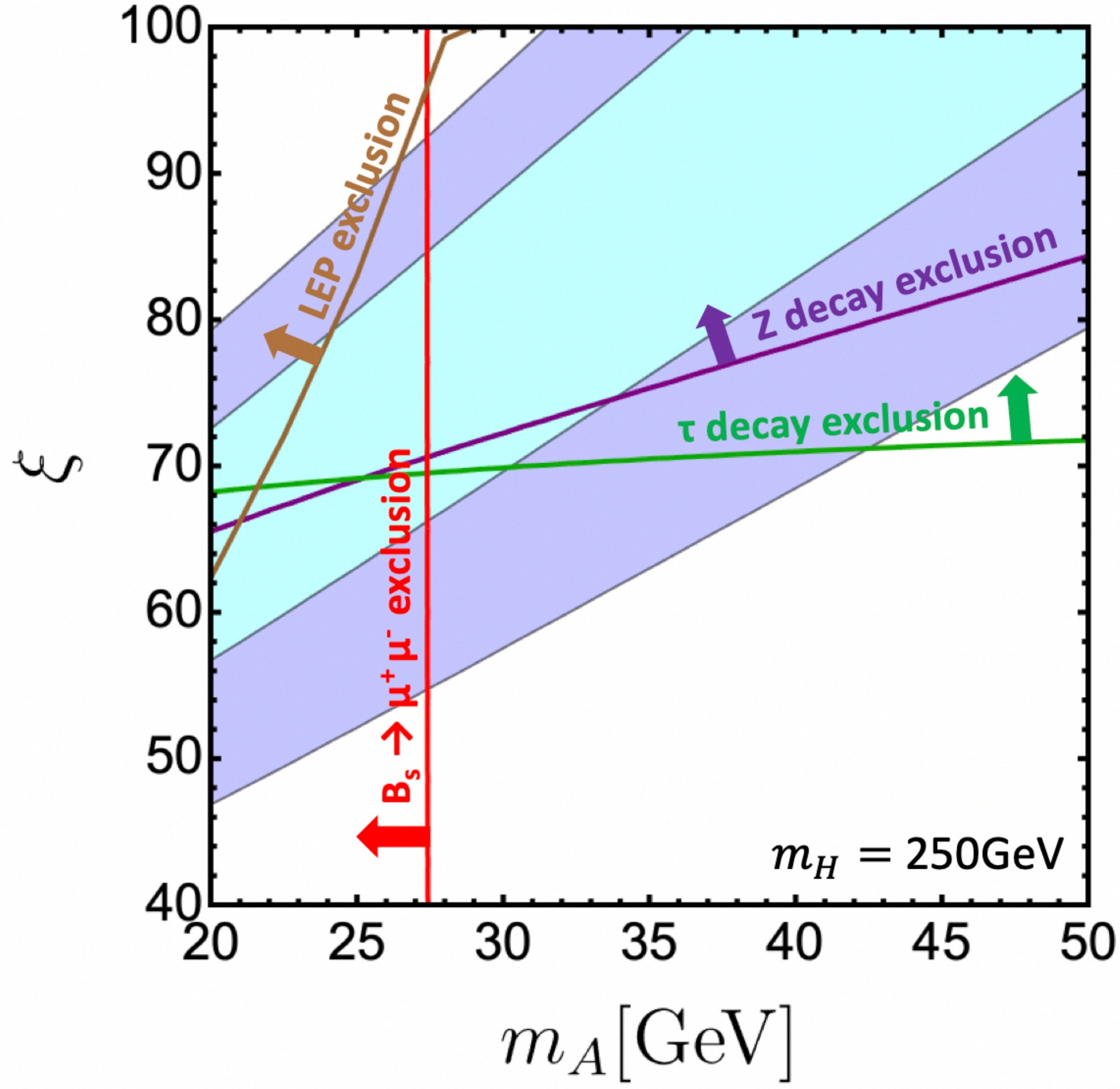} ~~~
\includegraphics[width=17em]{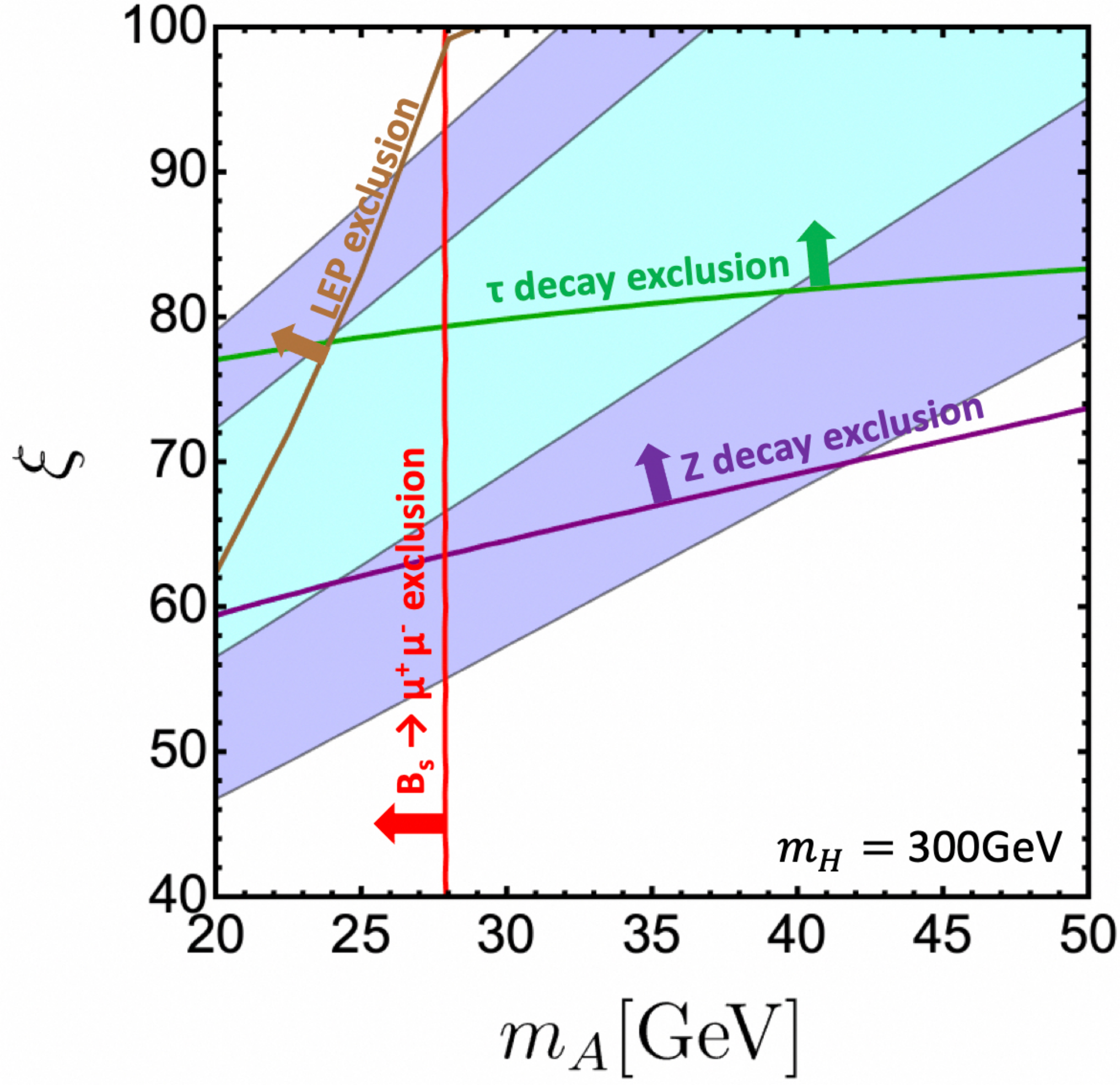}\\
\caption{
\label{Fig:Type-X}
The parameter plane of $\xi$ and $m_A$ in the type-X 2HDM.
The cyan and blue regions
can accommodate the muon $g-2$ anomaly at the 1$\,\sigma$ and 2$\,\sigma$ levels, respectively.
The region to the left of the green, purple, and brown lines is excluded by the $\tau$ decay, $Z$ decay, and scalar bremsstrahlung constraints, respectively.
The $B_s\to\mu^+ \mu^-$ constraint excludes the region to the left of the red line.
We take $m_H=m_{H^\pm}=250\,(300)\,$GeV on the left (right) panel.
} 
\end{center}
\end{figure}
In Fig.\,\ref{Fig:Type-X}, we show the parameter regions where the muon $g-2$ anomaly can be explained at the 1$\,\sigma$ and 2$\,\sigma$ level by the cyan and blue regions, respectively.
We take $m_H=m_{H^\pm}=250\,(300)\,$GeV on the left (right) panel.

In the regime of large lepton couplings in the type-X 2HDM, the tau (leptonic and hadronic) decays are modified by tree-level and one-loop corrections from the additional scalars.
The tree-level and one-loop corrections have been calculated in Refs.~\cite{Krawczyk:2004na,Abe:2015oca}.
Relevant formulae and the current experimental data are summarized in Appendix~\ref{app:tau_decay}.
Since the corrections from the additional scalars are suppressed by $1/m_{H^\pm}^2$, the $\tau$-decay bound uniformly becomes weaker, the heavier the additional scalar becomes.
Since the one-loop correction involving two tau-Yukawa couplings is larger than the one involving a single muon-Yukawa coupling, the box diagram is less important.
The excluded regions from tau decays are shown by the green lines in Fig.\,\ref{Fig:Type-X}.

In addition, the one-loop correction to the fermion coupling of the $Z$ boson provides an important cross-check in this scenario.
Thanks to the hierarchical structure of the Yukawa couplings, the $Z$-boson interaction with the tau leptons is most sensitive to the additional scalars.
For the partial cancellation of systematic uncertainties, taking the ratio of leptonic decay widths improves the sensitivity.
The LEP averages are given as \cite{ALEPH:2005ab},\footnote{It is noted that the recent result of $\frac{\Gamma(Z\to \mu^+\mu^-)}{\Gamma(Z\to e^+ e^-)}$ from ATLAS \cite{ATLAS:2016nqi}, which has a twice larger uncertainty, is consistent with the LEP result.
In addition, the uncertainty from LHCb is much larger than that of the LEP \cite{LHCb:2018ogb}.
Furthermore, the correlation matrix is not available in the PDG \cite{PDG2022}.
Therefore we use the LEP average.}
\begin{align}
\frac{\Gamma(Z\to \tau^+\tau^-)}{\Gamma(Z\to e^+ e^-)}=1.0019\pm0.0032\,, \quad
\frac{\Gamma(Z\to \mu^+\mu^-)}{\Gamma(Z\to e^+ e^-)}=1.0009\pm0.0028\,,
\end{align}
with a correlation of 0.63.
A larger mass difference of $m_H$--$m_A$ yields a larger deviation from the SM prediction of the $Z$-boson interaction.
Furthermore, the tau polarization asymmetry in $Z\to\tau^+\tau^-$ is also precisely measured at the LEP and the average is \cite{PDG2022}
\begin{align}
    A_\tau=0.143\pm0.004\,.
\end{align}
The corresponding corrections involving additional scalars are summarized in Appendix~\ref{app:Z_decay}.
The excluded regions from $Z$-boson decays are shown by the purple lines in Fig.\,\ref{Fig:Type-X}.
Contrary to the conventional decoupling behaviour, it is known that the additional scalar contributions are enhanced if $m_A \ll m_Z \ll m_H\simeq m_{H^\pm}$ is considered \cite{Chun:2016hzs}.
Therefore, the $Z$-boson bound becomes stricter for heavier additional scalars, which provides an exclusion region complementary to the $\tau$-decay bound.

Furthermore, the rare leptonic meson decay $B_s \to \mu^+ \mu^-$ gives a lower bound on $m_A$, which is independent of $\xi$, since in the diagram with a neutral scalar the $\xi$ dependence is cancelled.
In the past, the experimental world average of $\text{BR}(B_s \to \mu^+ \mu^-)$ had deviated from the SM prediction \cite{Bobeth:2013uxa,Beneke:2017vpq,Buras:2022wpw} by about $2\,\sigma$~\cite{CMS-PAS-BPH-20-003,Altmannshofer:2021qrr}.
However, the CMS collaboration recently reported the Run~2 full analysis and found the branching ratio to be consistent with the SM prediction \cite{CMS:2022dbz}.
As a result, the latest experimental world average is well consistent with the SM prediction \cite{HFLAV2023January}.
The dominant contribution to the $B_s\to \mu^+\mu^-$ comes from the light CP-odd scalar mediated diagram at one-loop, see Appendix~\ref{app:Bsmumu} for more details.
The new world average leads to a relaxed lower bound on $m_A$ compared to the previous world average.
We obtain $m_A \gtrsim 27$\,GeV in the type-X 2HDM, which is shown by the red lines in Fig.\,\ref{Fig:Type-X}.

The LEP probed electroweak $AH$ production in $4\tau$ final states ($e^+ e^- \to Z^\ast \to AH \to 2\tau^+ 2\tau^-$) without finding a significant excess over the SM expectation.
As a result a lower limit on the sum of the neutral scalar masses, $m_H+m_A \geq 190~(155)\,$GeV, has been obtained for the case where $\text{BR}(A\to\tau^+ \tau^-)\times\text{BR}(H\to\tau^+ \tau^-)=1 (0.1)$~\cite{DELPHI:2004bco,ALEPH:2006tnd}.
Lighter scalars have been also searched for in the scalar bremsstrahlung process ($e^+e^- \to \tau^+\tau^- A \to 2\tau^+ 2\tau$) \cite{DELPHI:2004bco}.
This constraint is especially stringent for a very light $A$.
An upper bound of $\xi \le34 ~(83)$ was obtained for $m_A=10~(25)\,$GeV with BR$(A \to\tau^+\tau^-)\simeq 1$.
Furthermore, there have been searches for a pair of $H^\pm$ in $\tau\nu$, $cs$ and $W^\pm A$ decay modes.
Again the absence of events exceeding the SM expectation allows us to set the lower mass limit as $ m_{H^\pm}\geq84$--$94\,$GeV, depending on $m_A $\cite{ALEPH:2013htx}.
The LEP exclusions, which come from the scalar bremsstrahlung process, are shown by the brown lines in Fig.\,\ref{Fig:Type-X}.

The LHC is also a powerful tool to search for additional scalars.
Due to the nature of a hadron collider, only partial information of a collision is accessible.
Thanks to the large statistics and good control of tau-lepton identification, low-mass charged Higgs bosons have recently been excluded~\cite{CMS:2018jrd,ATLAS:2019uhp}.
The Run~2 full result of the search for left-handed staus can be directly adapted to draw exclusion plots because the production cross section is the same as for charged-Higgs pair production.
The ATLAS measurement excluded $120\, {\rm GeV} \le m_{H^\pm}\le 390\, {\rm GeV}$ \cite{ATLAS:2019gti} while the CMS one excluded $115\, {\rm GeV}\le m_{H^\pm}\le 340\, {\rm GeV}$ \cite{CMS:2022rqk}, assuming BR$(H^\pm\to\tau\nu)=1$.
Since the SM background (BG) stems from $W^\pm$ pair production, it is difficult to probe lower values of $m_{H^\pm}$ at the LHC.
The low mass window $m_{H^\pm}\simeq100\,$GeV will close in the near future \cite{Iguro:2022tmr} once the systematic error scales as $1/\sqrt{L}$.
If a large mass difference between the neutral scalars is assumed, searches for same-sign $H^\pm$ become relevant \cite{Aiko:2019mww}.
However, a more dedicated experimental analysis is necessary to estimate the sensitivity.
It would be worth mentioning that the decay $h \to AZ$ is not possible in the alignment limit considered throughout this review.

In order to precisely interpret the constraints by the LHC searches, the decay properties of the additional scalars are important.
In addition to BR$(A\to\tau^+ \tau^-)\simeq1$, we summarize here the relevant parameter dependence as follows: 
\begin{align}
{\rm{BR}}(H \to \tau^+ \tau^-) & \simeq 
\frac{\Gamma(H \to \tau^+ \tau^-)}{\Gamma(H \to \tau^+ \tau^-) + \Gamma(H \to A Z)}\,,\\ 
{\rm{BR}}(H^\pm \to \tau\nu) & \simeq 
\frac{\Gamma(H^\pm \to \tau\nu)}{\Gamma(H^\pm \to \tau\nu) + \Gamma(H^\pm \to A W^\pm)}\,,\\
\Gamma(H \to \tau^+ \tau^-) & \simeq
\Gamma(H^\pm \to \tau \nu)=
\frac{m_{H}}{8\pi} \frac{m_\tau^2}{v^2} \xi^2\,,\\
\Gamma(H \to A Z)
& = \frac{m_{H}^3}{16\pi v^2}\lambda^{3/2}\left(\frac{m_A^2}{m_H^2},\frac{m_Z^2}{m_H^2}\right)\,,\\
\Gamma(H^\pm \to A W^\pm)
& = \frac{m_{H^\pm}^3}{16\pi v^2}\lambda^{3/2}\left(\frac{m_A^2}{m_{H^\pm}^2},\frac{m_W^2}{m_{H^\pm}^2}\right)\,,
\end{align}
where $\lambda(x_1,x_2)=(1-x_1-x_2)^2 - 4x_1x_2$ and $m_\tau$ is neglected in the phase-space factor.
Thus, for large $\xi$, tauonic scalar decays can make up a significant fraction of the total decay width.
For the relevant mass scale, $\xi \simeq 100$ yields branching ratios $\simeq {\cal O}(50\%)$, which means that the branching ratios and the resulting LHC constraints are sensitive in this interesting parameter region.

For the type-X 2HDM interpretation of the muon $g-2$ anomaly, the favored parameter regions are at large $\xi$ and very light $m_A \simeq 30$\,GeV, see Fig.\,\ref{Fig:Type-X}.
In this case, $\tau$-rich signatures at the LHC become relevant, as discussed in Ref.~\cite{Chun:2015hsa}.
The importance of the $\tau$-rich signatures is also discussed at the ILC~\cite{Chun:2019sjo}, where it is shown that the favored parameter regions will be fully covered.
Although a type-X specific search has not yet been performed, we can recast the current experimental searches for charginos and neutralinos in the MSSM, based on these $\tau$-rich signatures~\cite{CMS:2016gjw,ATLAS:2017qwn,ATLAS:2021moa,CMS:2021edw,ATLAS:2022nrb}.
In particular, we consider the ATLAS analysis \cite{ATLAS:2022nrb} which provides detailed kinematic cuts.
In the analysis, eight signal regions (SRs) are defined, designed such that the sensitivity to $\tilde\chi_1^+\tilde\chi_1^-$ and $\tilde\chi^\pm\tilde\chi^0_2$ events is enhanced.
As common features of all the SRs, at least two $\tau$-leptons, a veto on bottom quark jets in order to reject SM top-quark processes, and large stransverse mass \cite{Lester:1999tx,Barr:2003rg} $m_{\rm T2}>70$--$100$\,GeV are required.
Some of the SRs aiming for $\tilde\chi_1^+\tilde\chi_1^-$ and $\tilde\chi_1^+\tilde\chi_2^0$ require opposite-sign (OS) $\tau$ leptons, while the rest requires same-sign taus for $\tilde\chi_1^+\tilde\chi_2^0$.
As for the SRs with OS taus, a $Z/h$ veto ($m_{\tau\tau}>120$\,GeV) is required to capture the taus from stau decays, otherwise $m_{\tau\tau} \simeq m_h$ is required to capture the $h\to \tau^+ \tau^-$ in the decay chain.
Regarding the missing momentum, both possibilities of $E_{\rm miss}<150$\,GeV and $E_{\rm miss}>150$\,GeV are considered to capture the low-mass and the high-mass spectra, respectively.
Additional selection cuts are imposed depending on the SRs, for details see Ref.~\cite{ATLAS:2022nrb}.
As a result, $1$--$14$ events were observed in each SR for 139\,fb$^{-1}$ at $\sqrt{s}=13$~TeV, without significant excess.
This results in 95\% confidence level (CL) upper limits on the non-SM fiducial cross section ($\sigma_{\rm vis}^{95\%}$) of 0.03--0.1\,fb, which put severe bounds even on EW production processes.
This stringent constraint compared to the previous one is achieved by the very strong selection cut optimized for the chargino and neutralino searches and due to the large integrated luminosity of 139\,fb$^{-1}$.

\begin{figure}[t]
\begin{center}
\includegraphics[width=17.7em]{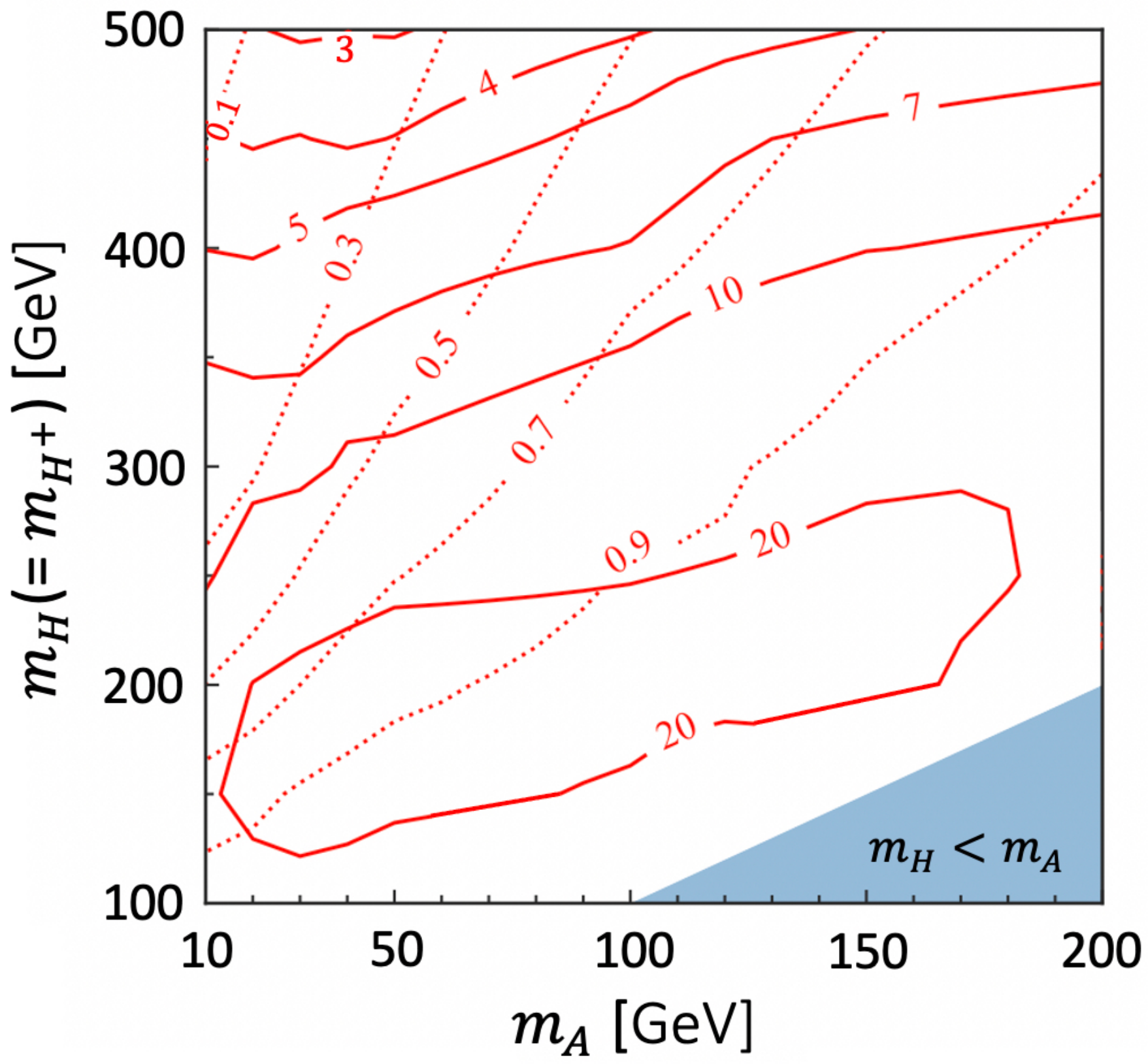} ~~
\includegraphics[width=17.7em]{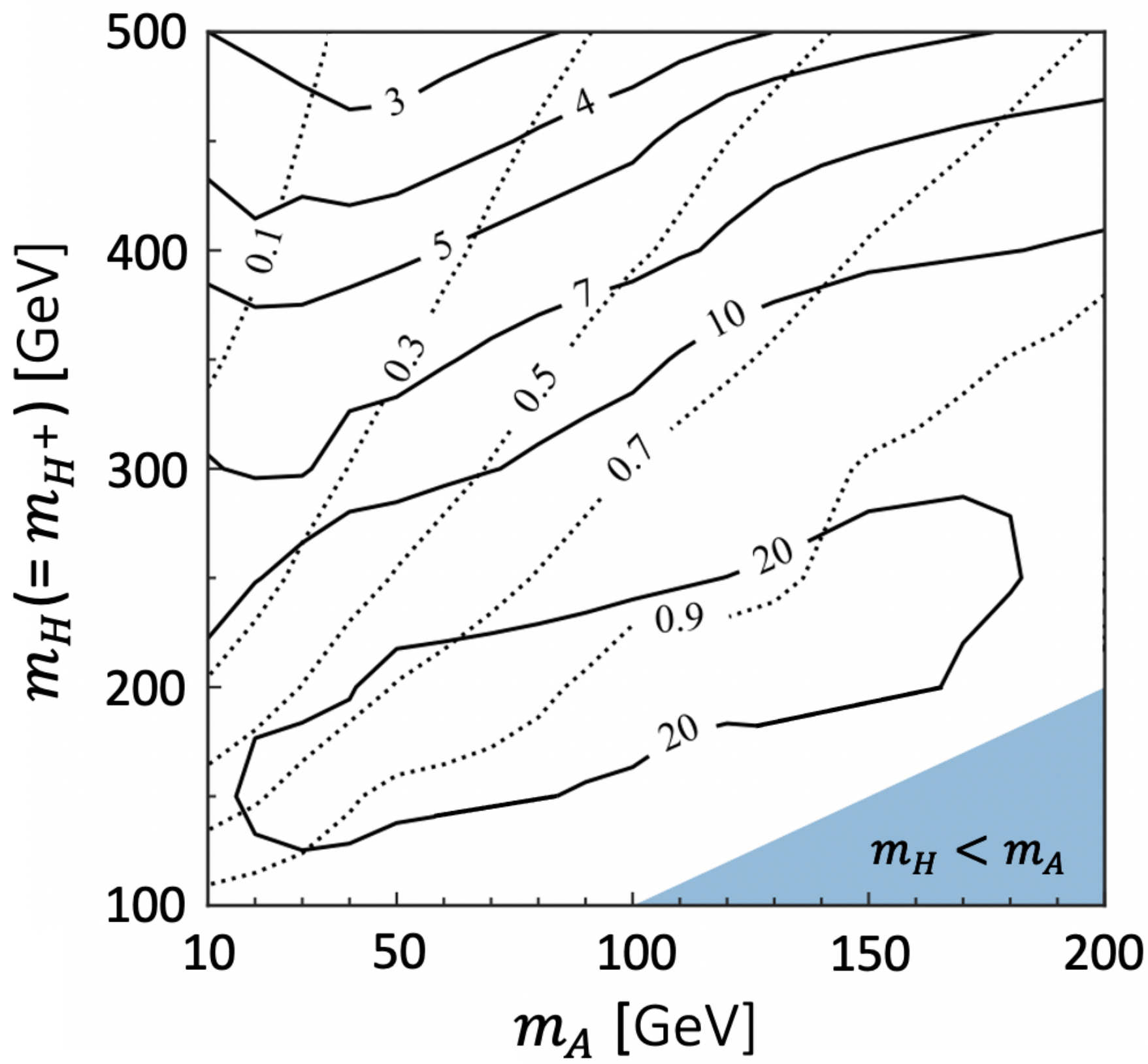}\\
\caption{
\label{Fig:Type-X_LHC}
The exclusion factor ($\sigma_{\rm vis}^{\rm type-X}/\sigma_{\rm vis}^{95\%}$) obtained by the MSSM $\tilde\chi^\pm \tilde\chi^0$ searches at the LHC and BR($H\to\tau^+ \tau^-$) are shown in the solid and dashed contours, respectively, in the type-X 2HDM.
The parameter $\xi$ is fixed to explain the muon $g-2$ anomaly at the $0\,\sigma$ (left panel) and $-2\,\sigma$ (right panel) levels, corresponding to $a_\mu^{\rm NP}=25.1\times 10^{-10}$ and $13.3\times 10^{-10}$, respectively.
The pale blue region in the bottom-right corner corresponds to $m_H<m_A$.
} 
\end{center}
\end{figure}

We generate EW pair-production events of additional scalars,
\begin{align}
p p \to HA,\,H^\pm A,\,H^\pm H,\,H^\pm H^\mp \,(\to \text{multi-}\tau)\,,
\end{align}
in the type-X 2HDM using {\sc\small MadGraph}5\_a{\sc\small MC}@{\sc\small NLO}~\cite{Alwall:2014hca} $+$ {\sc\small PYTHIA}~\cite{Skands:2014pea} $+$ {\sc\small DELPHES}~\cite{deFavereau:2013fsa} and apply the selection cuts defined for the above SRs \cite{ATLAS:2022nrb}.
We consider the cases where $m_A \le m_H=m_{H^{\pm}}$, and $10$\,GeV $ < m_A < 200$\,GeV, $100$\,GeV $< m_H < 500$\,GeV.
The total production cross section ranges from $5$\,fb to $4$\,pb.
For each model point, we generate 100K signal events.
Figure~\ref{Fig:Type-X_LHC} shows the contour plots of the ratio $\sigma_{\rm vis}^{\rm type-X}/\sigma_{\rm vis}^{95\%}$ (which we call exclusion factor), that is, how large an event number is expected relative to the 95\% CL upper limit.
The maximal exclusion factors are mainly from the SRs C1C1-LM and C1N2SS, which were designed to capture chargino pair-production events and chargino-neutralino pair-production events with same-sign $\tau$ signatures, respectively.
We only take the largest of these SRs to be conservative.
We fix $\xi\,(=\tan\beta)$ as to reproduce $\Delta a_\mu$ at the $0\,\sigma$ (left panel) and $-2\,\sigma$ levels (right panel), corresponding to $a_\mu^{\rm NP}=25.1\times 10^{-10}$ and $13.3\times 10^{-10}$, respectively.
The contours in dotted lines show the expected value of BR$(H \to \tau^+ \tau^-)$.
The results show that the interesting regions for the muon $g-2$ anomaly are completely excluded and it is difficult to save this model unless new decay modes are introduced.
We also checked that selecting any values of ${\rm{BR}}(H\to \tau^+ \tau^-)$ and ${\rm{BR}
}(H^\pm\to \tau\nu)$
results in the whole region of the depicted plane being excluded as long as 
${\rm{BR}}(H\to \tau^+ \tau^-)+ {\rm{BR}}(H\to ZA)= 1$ and 
${\rm{BR}}(H^\pm \to \tau\nu) + {\rm{BR}}(H^+\to W^+A) =1$.
Even if setting ${\rm BR}(H\to \tau^+ \tau^-)={\rm BR}(H^\pm\to \tau \nu)=0$, the lowest value of the exclusion factor is about 1.6.
These results are the updated plots of Fig.\,7 in Ref.~\cite{Chun:2015hsa}.

\begin{figure}[t]
\begin{center}
\includegraphics[width=20em]{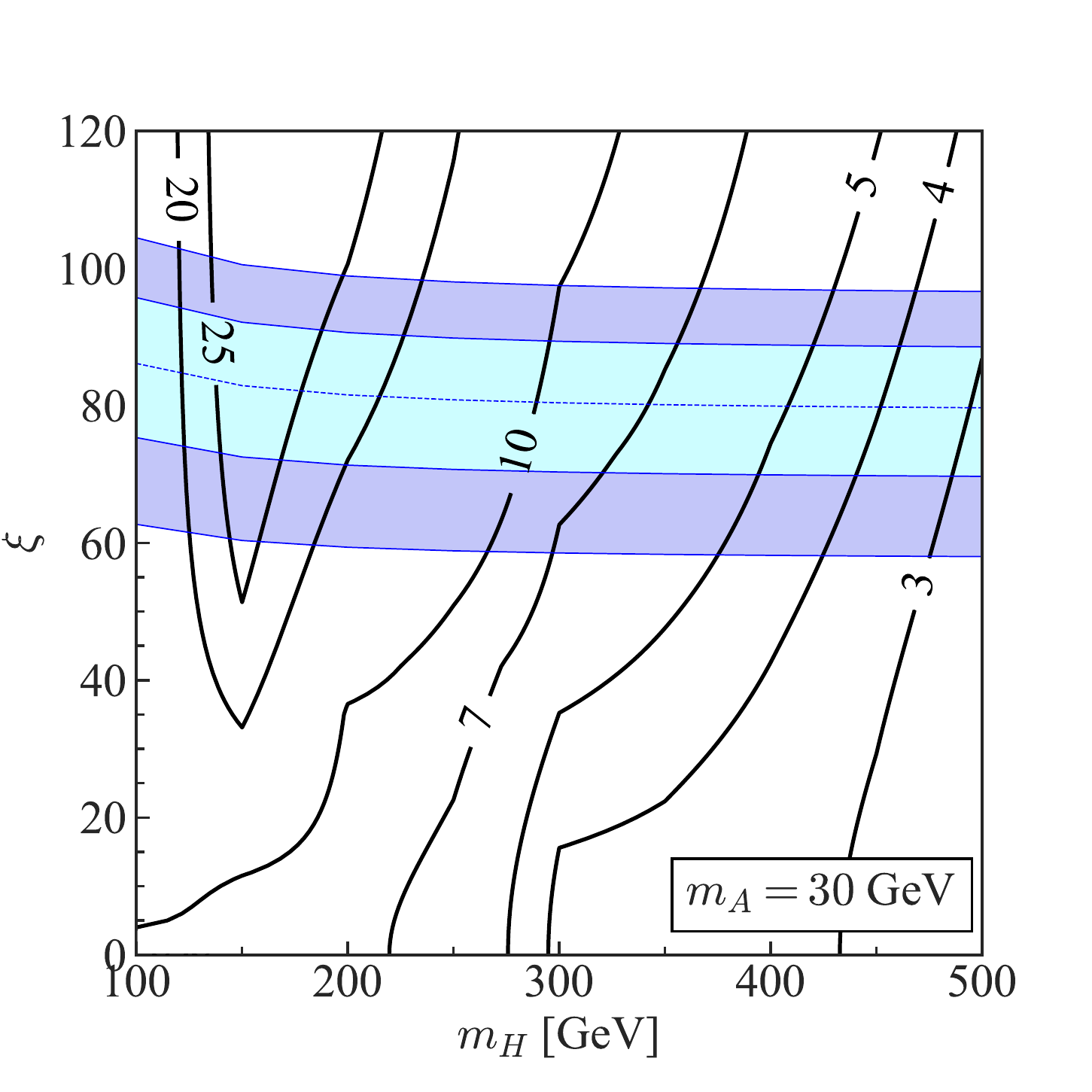} 
\caption{
\label{Fig:Type-X_LHC_mA30}
The contour plot of the exclusion factor ($\sigma_{\rm vis}^{\rm type-X}/\sigma_{\rm vis}^{95\%}$) with fixed $m_A=30$\,GeV, in the type-X 2HDM.
The cyan and blue regions can explain the muon $g-2$ anomaly at the $1\,\sigma$ and $2\,\sigma$ levels, respectively.
} 
\end{center}
\end{figure}
Finally, we show the same exclusion factor defined above in the $\xi$--$m_H(=m_{H^\pm})$ plane with fixed $m_A = 30\,$GeV in Fig.\,\ref{Fig:Type-X_LHC_mA30}.
We note that the figure does not change drastically even if we set $m_A = 50\,$GeV as one can infer from contours of the exclusion factor in Fig.\,\ref{Fig:Type-X_LHC}.
In general, if $(m_H - m_A) > m_Z,m_W$, the exclusion factor becomes larger as the ratio $\xi / m_H$ increases.
This is due to the larger $\tau$-branching ratio where the acceptance in the relevant SRs is larger in the tau modes.
If $(m_H - m_A) < m_Z,m_W$, specifically at $m_H=100$\,GeV in this plot, the exclusion factor becomes smaller although still larger than ten.
The cyan and blue regions can explain the muon $g-2$ anomaly at the $1\,\sigma$ and $2\,\sigma$ levels, respectively.
One can also see explicitly that even for ${\rm{BR}}(H\to \tau^+\tau^-)={\rm{BR}}(H^\pm\to \tau\nu)=0$, which is equivalent to the $\xi \to 0$ limit, the exclusion factor is always more than 1, with the minimum around 2 in the bottom-right corner.

In conclusion, the type-X 2HDM interpretation of the muon $g-2$ anomaly in the mass regions we considered is completely excluded by the chargino pair-production and chargino-neutralino pair-production searches at the LHC Run~2, as shown in Figs.~\ref{Fig:Type-X_LHC} and \ref{Fig:Type-X_LHC_mA30}.
It is impressive that these searches using the tremendous luminosity accumulated at the LHC already provide such high sensitivities for EW production processes.

\subsection{Flavor-aligned 2HDM}
\label{sec:fa_2HDM_pheno}

In the FA2HDM, the Yukawa couplings $\rho_f$ are proportional to the mass matrix, but the overall constants of up-type and down-type quarks and leptons are independent free parameters.
Therefore, the model includes the type-X 2HDM and thus has a broader parameter space.

In addition to the Barr-Zee diagram with the tau-lepton in the loop, the one with a top quark in the loop could contribute to the muon $g-2$.
However, the constraint from $B_s\to \mu^+\mu^-$ sets an upper limit on the top-loop contribution to the muon $g-2$.
This implies a stringent constraint for the light CP-odd scalar scenario.
It is worth mentioning that $B_s \to \mu^+ \mu^-$ can also receive ``type-II 2HDM''-like contributions proportional to $\xi_d\,\xi_e$.
If $\xi_d$ is of order $\xi_u m_t / m_b$, these contributions can become significant despite the much lighter bottom-quark mass.
This can weaken the top-quark loop constraints from $B_s \to \mu^+ \mu^-$, in particular if a fine-tuned cancellation between $\xi_d$ and $\xi_u$ occurs.
However, we do not consider such cancellations here and thus neglect contributions involving $\xi_d$.

\begin{figure}[t]
\begin{center}
\includegraphics[width=17em]{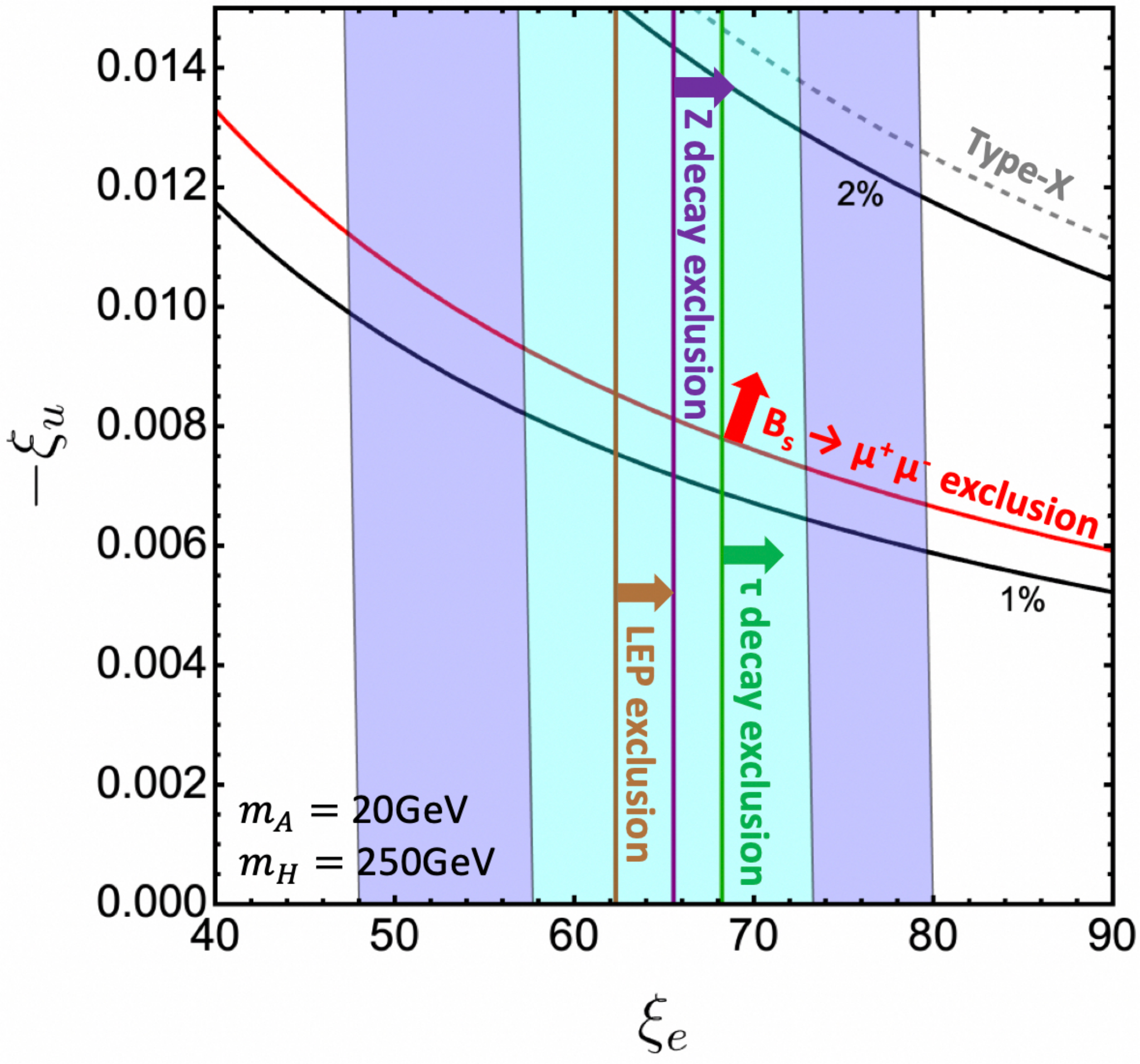} ~~~
\includegraphics[width=17.1em]{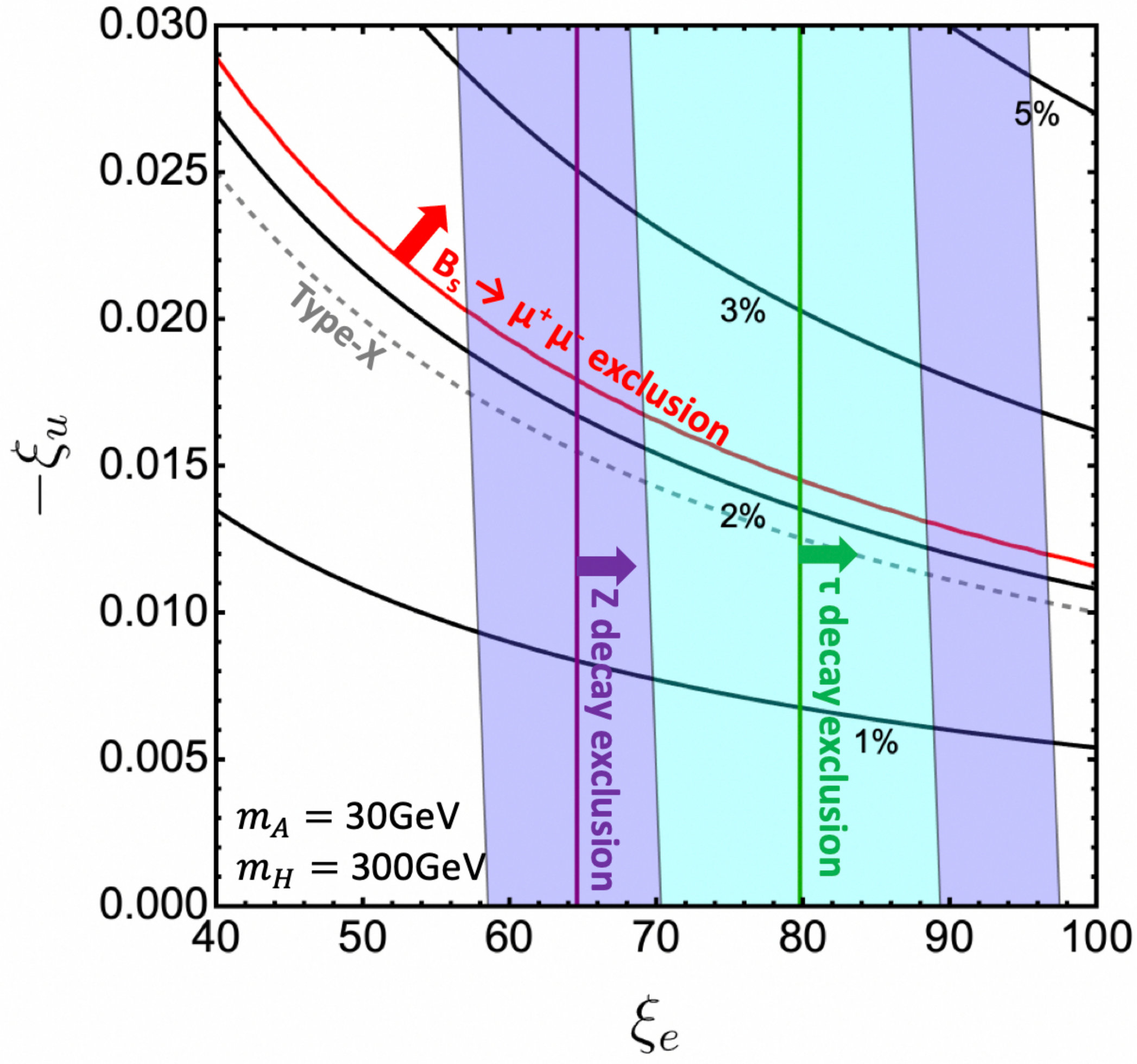}\\
\caption{
\label{Fig:FA}
The parameter plane spanned by $\xi_e$ and $\xi_u$ in the FA2HDM.
The cyan and blue regions accommodate the muon $g-2$ anomaly at the $1\,\sigma$ and $2\,\sigma$ levels, respectively.
The regions above the red lines are excluded by $B_s \to \mu^+ \mu^-$ at the 95\% CL.
The black contours show the size of the top-loop Barr-Zee contribution in units of $\Delta a_\mu$.
The regions to the right of the green, purple and brown lines are excluded by $\tau$ decay, $Z$ decay, and scalar bremsstrahlung constraints, respectively.
We take $m_A=20\,(30)\,$GeV and $m_H=m_{H^\pm}=250\,(300)\,$GeV on the left (right) panel.
The gray dashed lines represent the type-X 2HDM limit.
} 
\end{center}
\end{figure}
In Fig.\,\ref{Fig:FA}, we show the muon $g-2$ favored region in the $\xi_e$ vs.~$\xi_u$ plane where $m_A=20\,$GeV and $m_H=m_{H^\pm}=250\,$GeV, and $m_A=30\,$GeV and $m_H=m_{H^\pm}=300\,$GeV are fixed on the left and right panels, respectively.
The red, green, purple and brown lines are excluded by $B_s \to \mu^+ \mu^-$, $\tau$ decay, $Z$ decay, and scalar bremsstrahlung constraints, respectively.
The black contours show the size of the top-loop Barr-Zee contribution in units of $\Delta a_\mu$.
It is found that the top-quark loop constitutes up to about $1\%$ of the total deviation for $m_A=20\,$GeV.
For $m_A=30\,$GeV, the contribution is still less than $2\%$.
The constraint becomes more stringent when $A$ gets lighter.
We found that even with $m_A=50\,$GeV the contribution can only be up to $5\,\%$.
In order to evade the stringent bound from $B_s \to \mu^+ \mu^-$, $\xi_u$ needs to be as small as $\mathcal{O}(0.01)$, implying that this scenario is almost identical to the type-X 2HDM in the context of the muon $g-2$ anomaly.
Note that the gray dashed lines in Fig.\,\ref{Fig:FA} represent the type-X 2HDM limit.

If the Yukawa coupling $\xi_u$ is small, other flavor observables and collider processes are less affected, \eg, the single-scalar production via gluon fusion and di-Higgs production cross section are too small to be probed \cite{Iguro:2022fel}.
Therefore, the key probe is the same as in the type-X 2HDM case and we refer the reader to the discussion in the previous section.
As a result, multi-$\tau$ signatures at the LHC exclude the explanation of the muon $g-2$ anomaly within the FA2HDM.
We emphasize that the presence of $\xi_d$ cannot significantly reduce the $A\to \tau^+ \tau^-$ branching ratio since $\xi_e$ is already large in any attempt to explain the muon $g-2$ anomaly.
However, the coupling $\xi_d$ could also contribute to the production of scalars and $B_s\to\mu^+ \mu^-$.

\subsection{Muon-specific 2HDM}
\label{sec:model_m2HDM_pheno}

\begin{figure}[p]
\begin{center}
\includegraphics[width=21em]{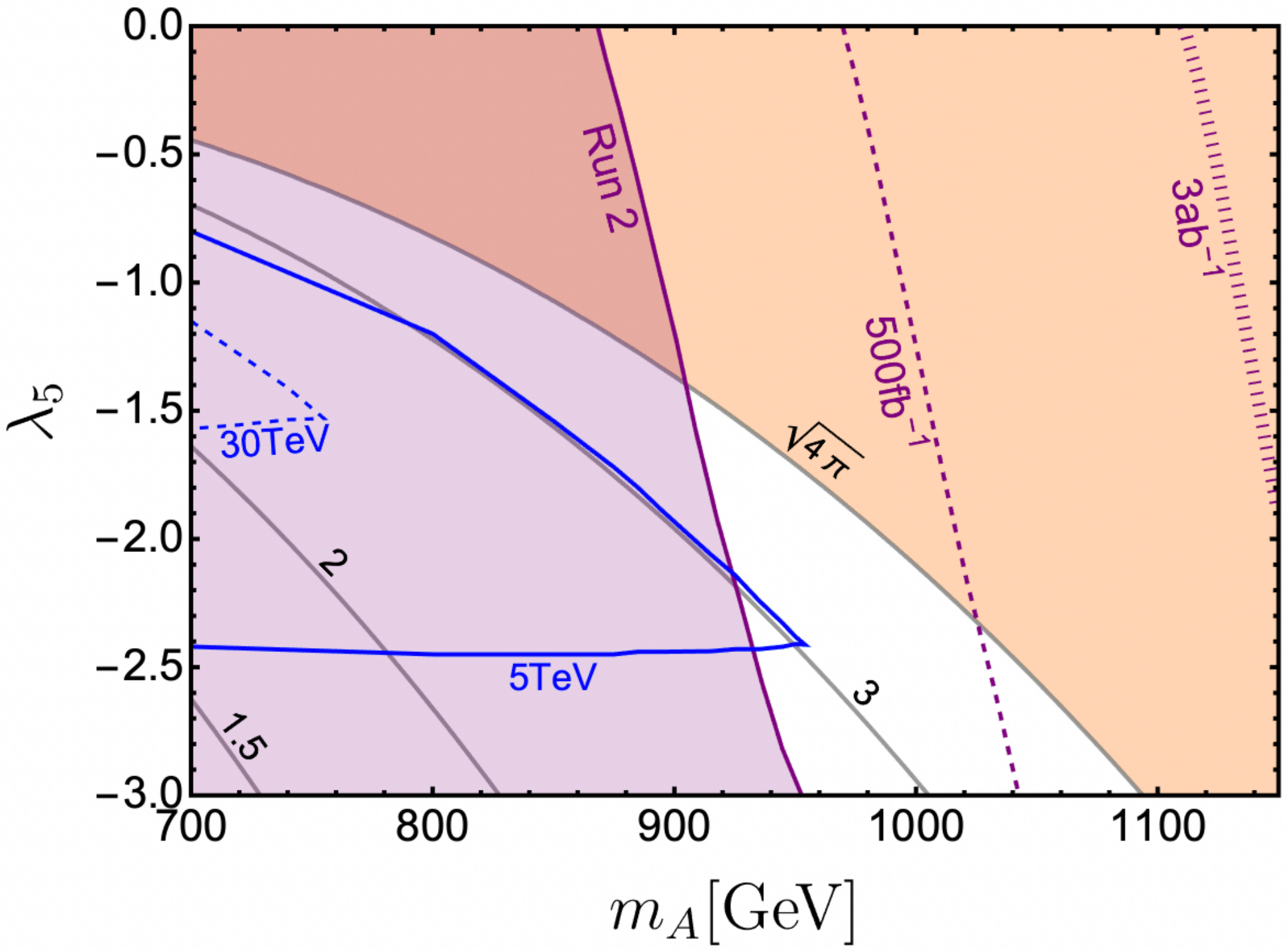}\\[0.3cm]
\includegraphics[width=21em]{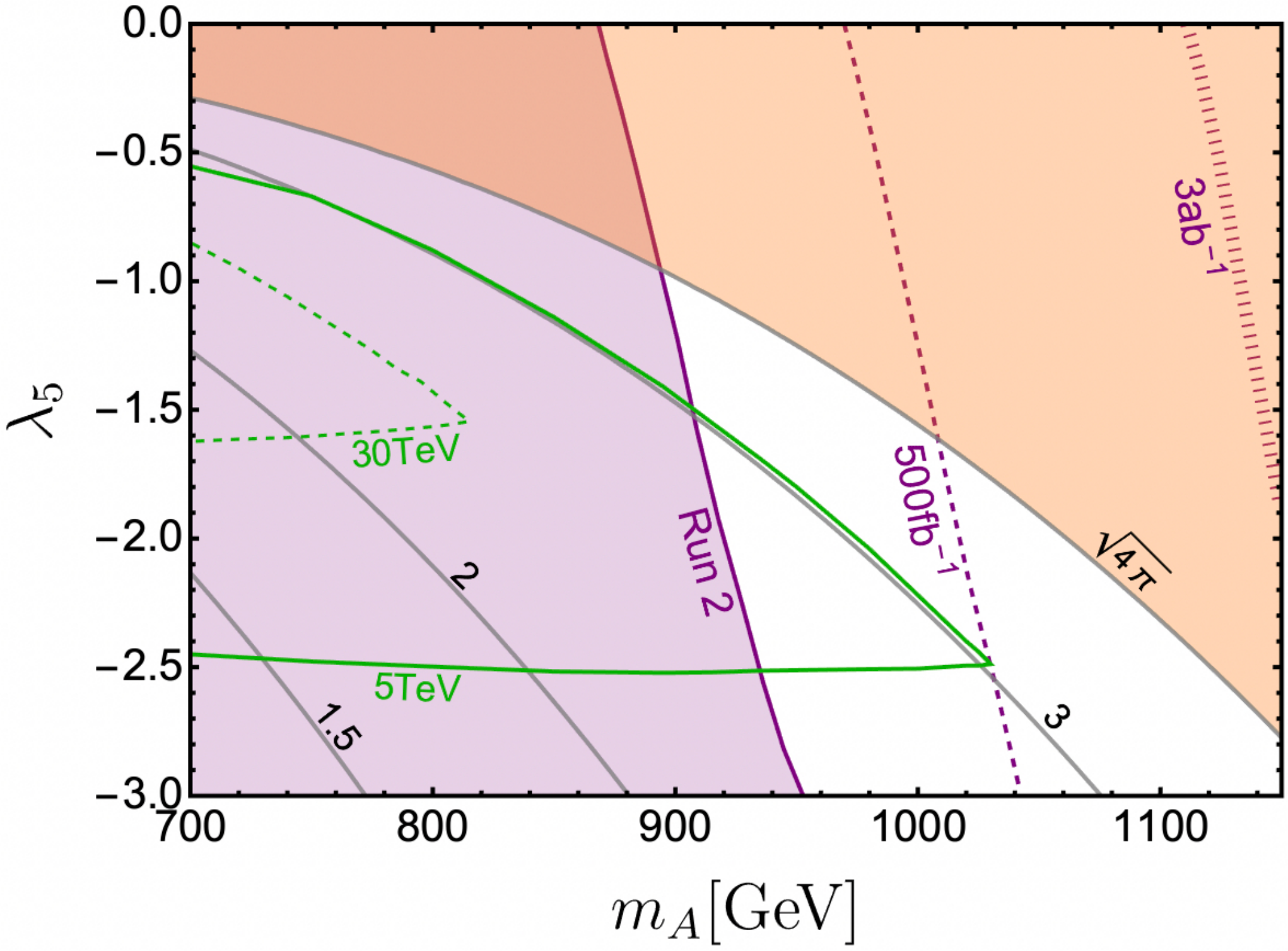}\\[0.3cm]
\includegraphics[width=21em]{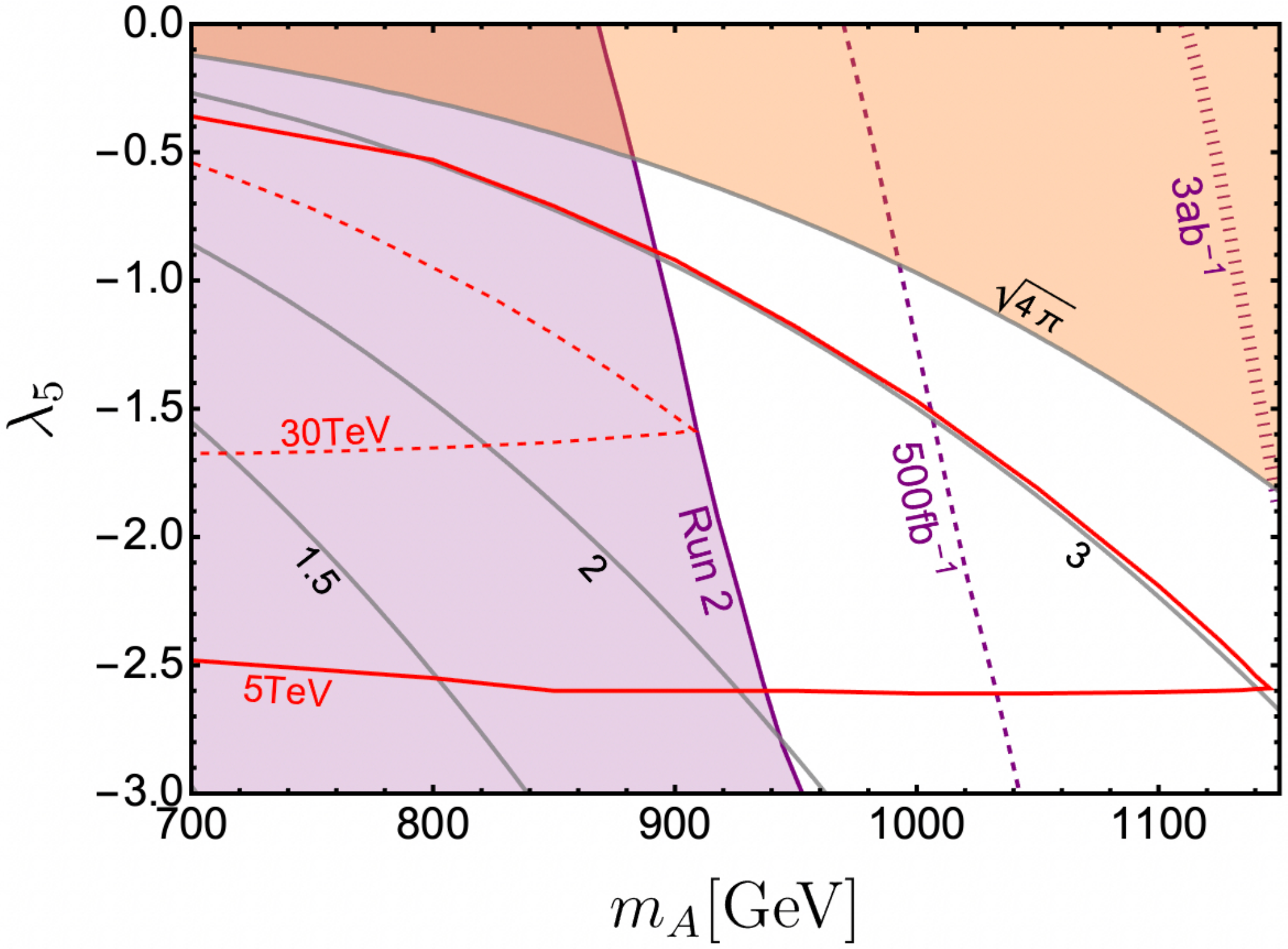} 
\caption{
\label{Fig:mu2HDM_summary}
The $\lambda_5$--$m_A$ parameter plane in the $\mu$2HDM.
The black contours correspond to the value of $\rho_e^{\mu\mu}$ that explains the muon $g-2 $ anomaly at the $0\,\sigma$ (upper), $-1\,\sigma$ (middle), and $-2\,\sigma$ (bottom) levels.
The orange regions violate the perturbative unitarity bound.
The purple regions are excluded by the high-$p_T$ multi-lepton searches based on our simulation.
The dashed and dotted purple lines correspond to the extended future prospect with an integrated luminosity of 500\,fb$^{-1}$ and 3\,ab$^{-1}$.
The solid and dashed colored (blue, green, and red) lines in each panel correspond to the parameter regions where the cutoff scale are $5\,$TeV and $30\,$TeV, respectively.
} 
\end{center}
\end{figure}

In the $\mu$2HDM, the one-loop contribution to $a_\mu^{\rm NP}$ can explain the muon $g-2$ anomaly, while the two-loop Barr-Zee contribution is suppressed by the mass of the heavy additional scalars and the electromagnetic coupling constant $\alpha$.
It is known that $H$ should be lighter than $A$ in order to have the correct sign of $a_\mu^{\rm NP}$.
In this case $m_H\simeq m_{H^\pm}$ is favored to satisfy both the vacuum stability condition and $T$-parameter constraint \cite{Gerard:2007kn}.
The size of the mass difference is controlled by the Higgs quartic couplings and thus constrained by the RGE-based perturbative unitarity constraint, see Appendices \ref{app:PUnVS} and \ref{app:beta} for details.
The black lines of Fig.~\ref{Fig:mu2HDM_summary} show the values of $\rho_e^{\mu\mu}$ required to explain the central value of the muon $g-2$ anomaly (top), $-1\,\sigma$ (middle), and $-2\,\sigma$ (bottom) levels, 
corresponding to $a_\mu^{\rm NP}=25.1\times 10^{-10}$, $19.2\times 10^{-10}$, and $13.3\times 10^{-10}$, respectively.
The orange region requires $\rho_e^{\mu\mu}\ge \sqrt{4\pi}$ (corresponding to $\xi_e \simeq 5900$), violating perturbativity, and thus we do not consider it.

Since the additional scalars mainly couple to muons, direct searches at the LHC, \eg, searches for smuons or multi-lepton final states, give a lower bound on the masses of the additional scalars.
Previously, the authors of Ref.~\cite{Abe:2017jqo} found that the multi-lepton search performed with the 35.9\,fb$^{-1}$ data set at $\sqrt{s}=13$\,TeV \cite{CMS:2017wua} excludes Higgs masses below $m_H\simeq620$\,GeV.\footnote{$m_A=m_{H^\pm}=m_H+90$\,GeV is assumed to derive the bound.}
We updated the analysis with the Run\,2 full data \cite{CMS:2019xud}.
Note that there is a similar search for NP in multi-lepton final states~\cite{ATLAS:2020wop}, originally motivated by heavy vector-like leptons in the type-III seesaw model.
However, we cannot directly use the result since additional jets are required.

\begin{figure}[t]
    \begin{center}
\includegraphics[width=12em]{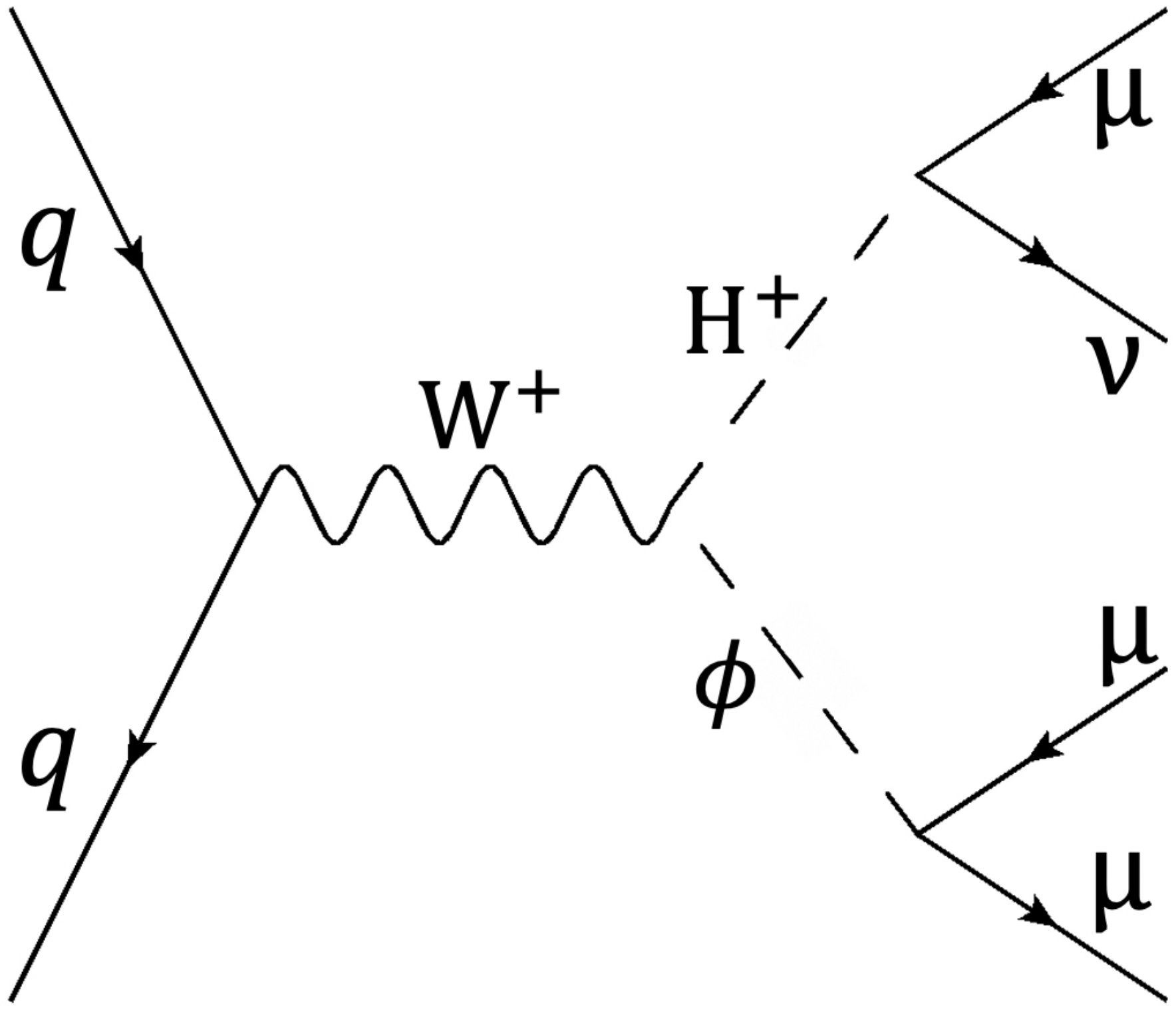} ~~~~~~~~~
\includegraphics[width=12em]{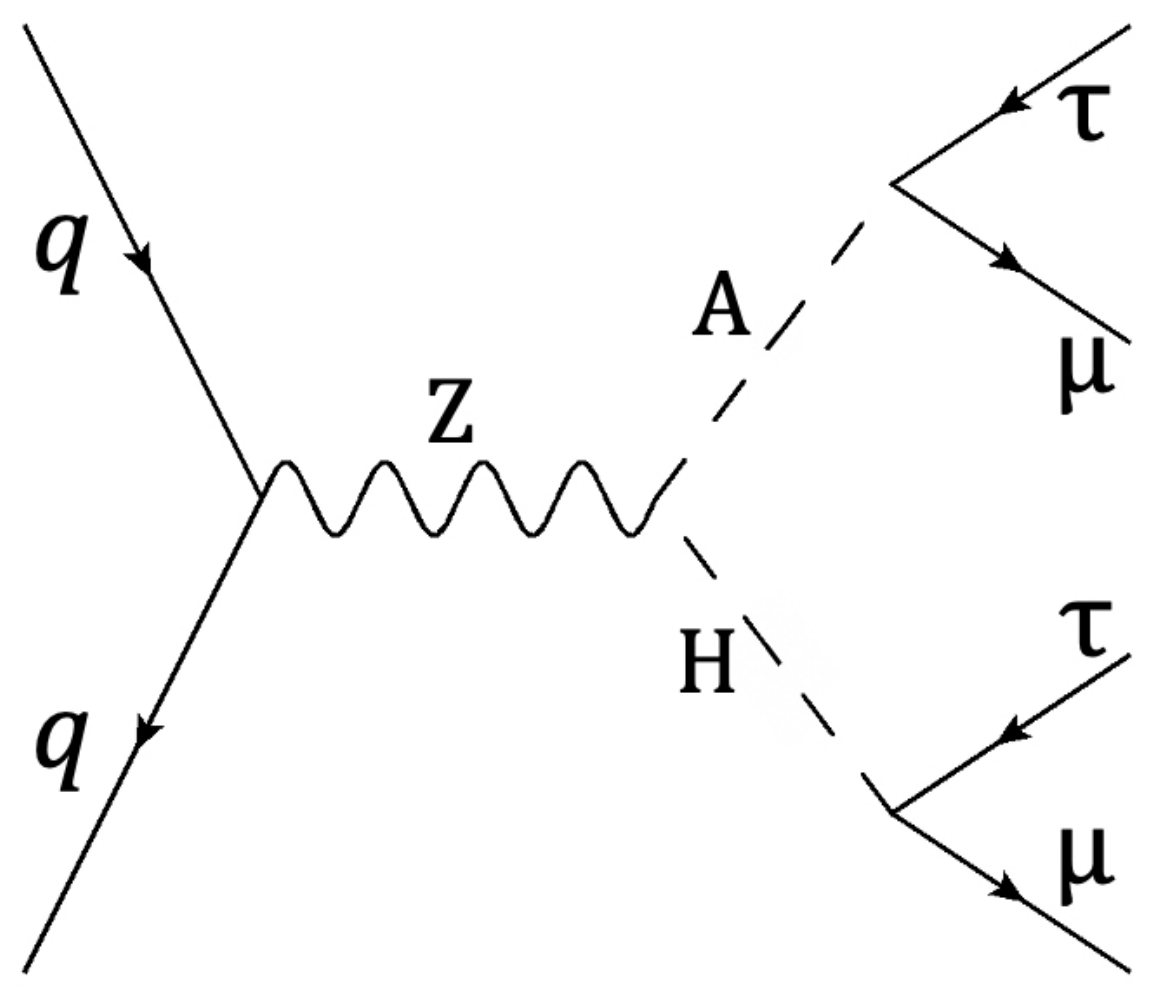} \\
\hspace{30em}
\includegraphics[width=12em]{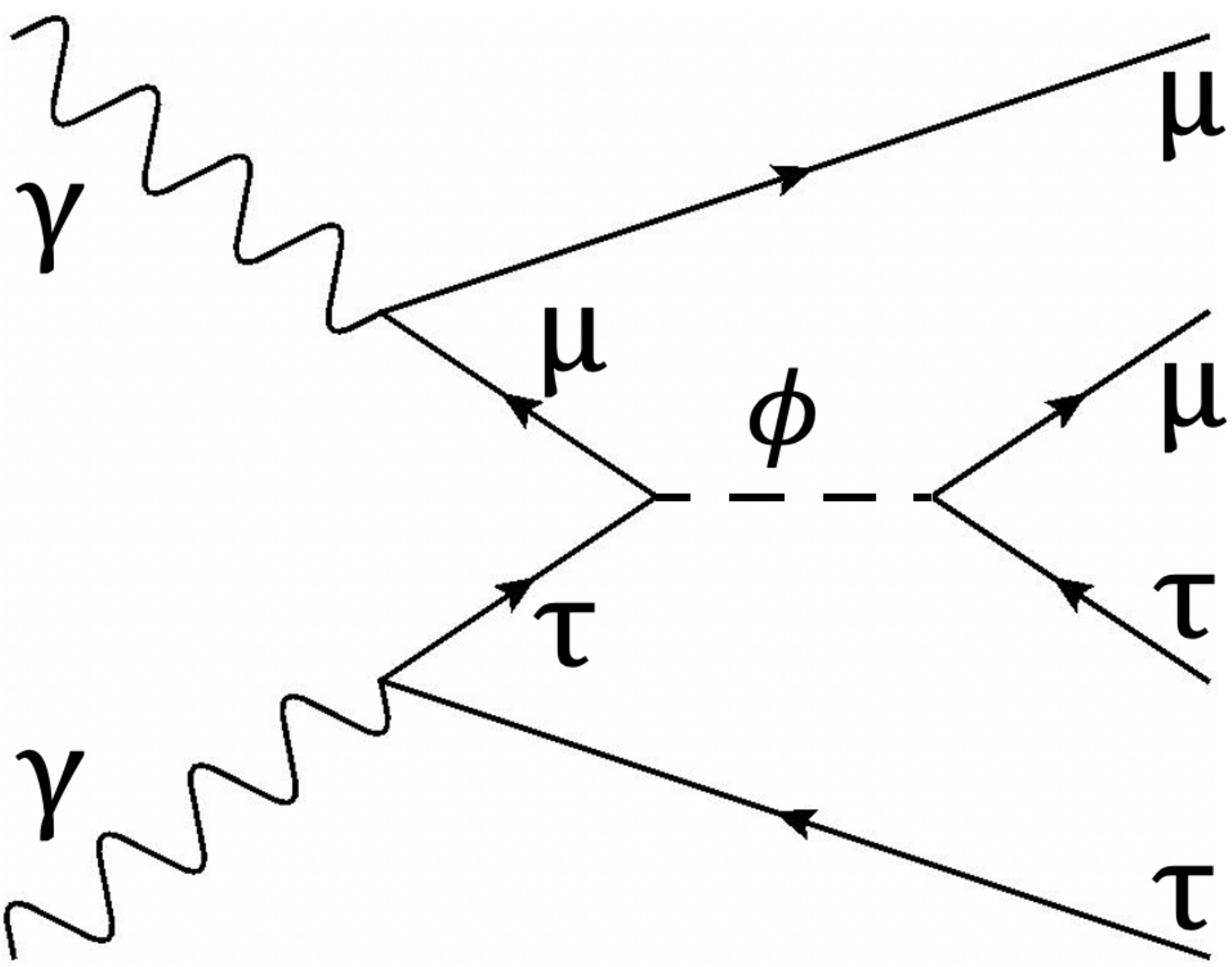} ~~~~~~~~~
\includegraphics[width=12em]{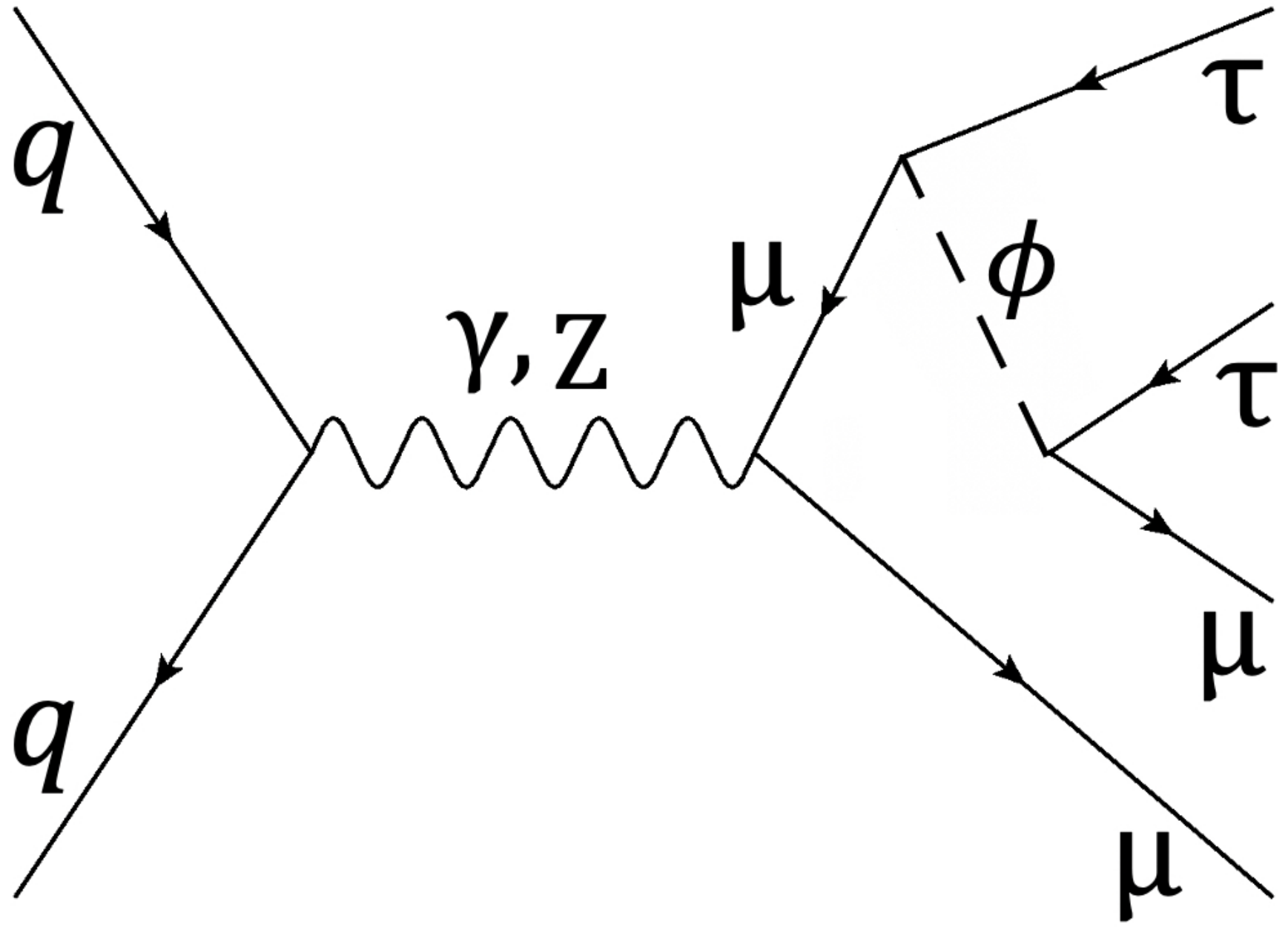} \hspace{10em}
\caption{
\label{Fig:diaLHC}
Representative Feynman diagrams for the collider searches in the $\mu$2HDM (top-left) and $\mu\tau$2HDM (the others).} 
\end{center}
\end{figure}

We generated 100K signal events for the process $pp\to \phi H^\pm\to 3\mu+\nu_\mu$, shown in Fig.\,\ref{Fig:diaLHC} (top-left), with {\sc\small MadGraph}5\_a{\sc\small MC}@{\sc\small NLO}~\cite{Alwall:2014hca} for a given set of $H^\pm$ and $\phi$ masses at $\sqrt{s}=13\,$TeV, where $\phi$ denotes $A$ or $H$.
Then the scalar sum of charged-lepton $p_T$ (L$_T$) and missing transverse energy (MET) is calculated and compared to the result in Fig.\,3-c of Ref.~\cite{CMS:2019xud} where both muon and electron are considered.
If muon-exclusive data become available, the signal-to-background ratio will be amplified, yielding improved sensitivity.
We evaluated the Run~2 exclusion region from the multi-lepton search,
which is depicted by the purple shaded region in Fig.\,\ref{Fig:mu2HDM_summary}.
Compared to the previous study \cite{Abe:2017jqo}, the lower mass limit is increased by about 200\,GeV.

The future sensitivity is estimated by assuming the significance scales as the square root of the luminosity.\footnote{Since the high-$p_T$ lepton signal region is currently statistically limited, this treatment is justified.} 
We point out that better sensitivity would be obtained with larger L$_{\rm{T}}$+MET bins; however, this would entail a more complicated experimental analysis.
Therefore, our procedure gives a conservative estimate regarding sensitivity.
We note that smuon searches at the LHC give less stringent constraints \cite{CMS:2018eqb,Aaboud:2017yvp}.

Furthermore, the Landau pole scale $\Lambda_{\rm{LP}}$ is shown as a colored contour in Fig.\,\ref{Fig:mu2HDM_summary}, where we use solid and dashed lines to illustrate $\Lambda_{\rm{LP}}=5\,$TeV and $30\,$TeV, respectively.
There is still a small region that can explain the central value of the muon $g-2$ anomaly if one requires the theory to be perturbative at $5\,$TeV.
On the other hand, if $\Lambda_{\rm{LP}}\ge30\,$TeV is required, the model cannot explain the muon $g-2$ anomaly in any region of the parameter space.

We stress that a future 500\,fb$^{-1}$ data set, which approximately corresponds to the integrated luminosity of the Run\,3 full data, can probe the whole $1\,\sigma$ region once we require that perturbative theory holds at least up to 5\,TeV.
The current $2\,\sigma$ region can be covered with a data set of 3\,ab$^{-1}$.

It is noted that our analysis can be readily applicable to the flavor conserving scenario of general two Higgs doublet model discussed in Refs.\,\cite{Botella:2020xzf,Botella:2022rte}.

\subsection{\texorpdfstring{$\mu\tau$}{mutau}-flavor violating 2HDM}
\label{sec:mutau_2HDM_pheno}

In the $\mu\tau$2HDM, the dominant contribution to $a_{\mu}^{\rm NP}$ arises from the one-loop diagram involving a scalar and a tau lepton, see Fig.\,\ref{Fig:diag-2} (right).
The contribution receives an $m_\tau / m_\mu$ enhancement factor compared to the $\mu$2HDM, due to the chirality flip on the internal tau propagator.
Thanks to this enhancement, heavier scalars can serve as an explanation for the muon $g-2$ anomaly, in contrast to the $\mu$2HDM.

Again the mass difference $m_H$--$m_A$ needs to be large to explain the muon \mbox{$g-2$}, see Appendix~\ref{app:muong-2}.
Note that the mass difference is determined by $\lambda_5$ in the Higgs alignment limit, see \eq{Eq:mass_l5}.
This implies large scalar couplings in the Higgs potential, and thus the RGEs become important, see Appendices \ref{app:PUnVS} and \ref{app:beta} for details.
Once we require that the model remains perturbative up to 30\,TeV (5\,TeV), we obtain an upper limit on the scalar mass scale of 1250\,GeV (1650\,GeV).

For this model, direct searches at the LHC can effectively put constraints on the available parameter space.
The additional scalars are quark-phobic (see \eq{eq:mutaurho}) and thus the main production mechanism at hadron colliders is EW pair production.
The heavy neutral scalar dominantly decays into $\mu\tau$.
The unique double $\mu\tau$-flavor violating resonances can result in two same-sign muons and two oppositely charged same-sign tau leptons in the final state, shown in Fig.\,\ref{Fig:diaLHC} (top-right), while the SM background can be neglected to a good approximation.
There is so far no experimental analysis for this channel, however theoretical sensitivity studies are available \cite{Iguro:2019sly,Blanke:2022kpi}.
In these studies, single-scalar production diagrams via Yukawa interactions are taken into account, shown in Fig.\,\ref{Fig:diaLHC} (bottom diagrams), in addition to EW pair production, enhancing the sensitivity in the scenario of large masses.

\begin{figure}[t]
\begin{center}
\includegraphics[width=29em]{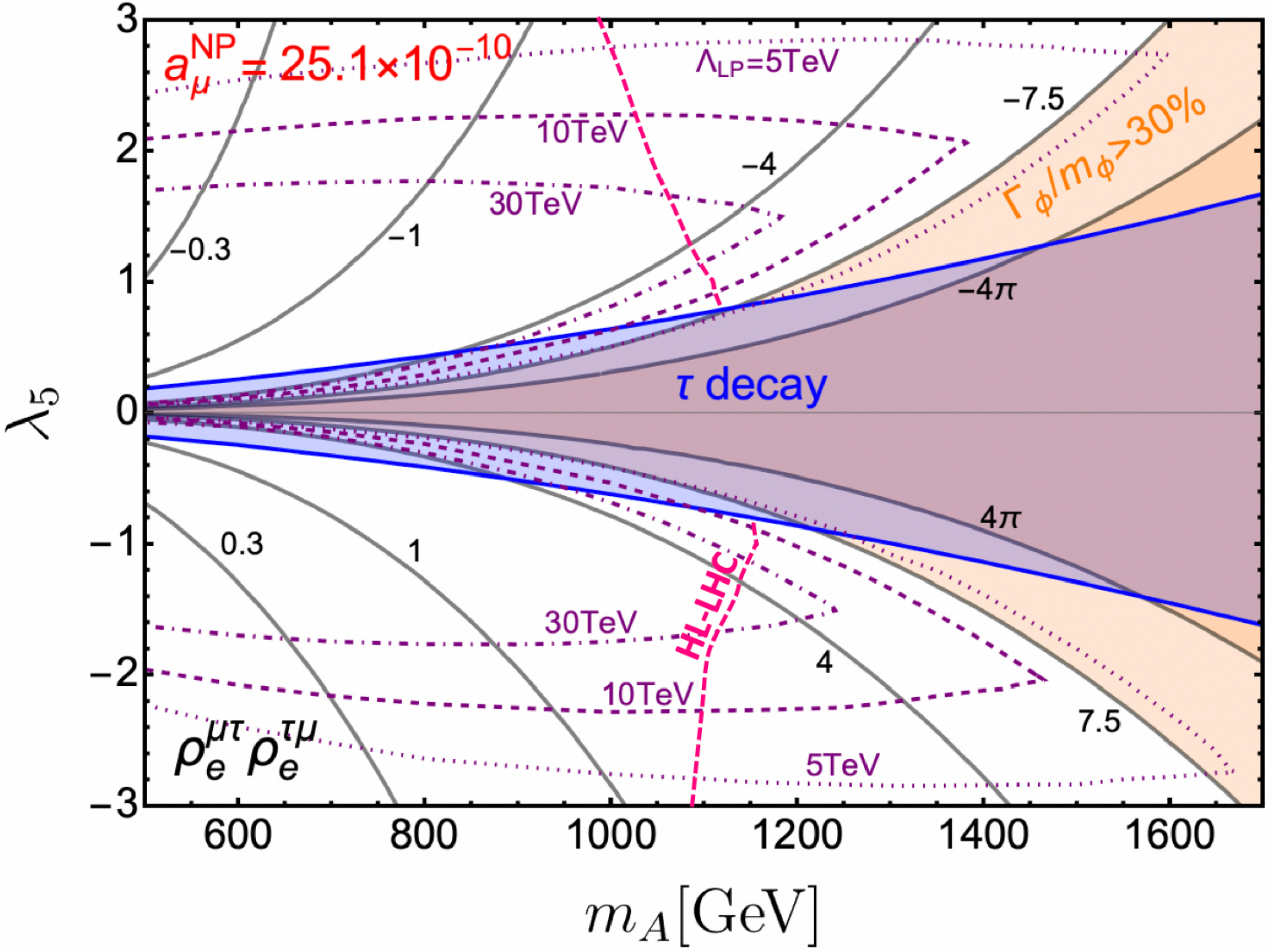}
\caption{
\label{Fig:mutau2HDM_summary}
The $\lambda_5$--$m_A$ parameter plane in the $\mu\tau$2HDM.
The black contour corresponds to the value of $\rho_e^{\mu\tau}\rho_e^{\tau\mu}$ that explains $a_\mu^{\rm{NP}}=25.1\times10^{-10}$.
The cutoff scale is depicted by the purple contour.
The blue shaded region is excluded by the tau decay constraint.
The dashed magenta line corresponds to the future prospect of the HL-LHC, the region to the left of which will be covered.
A future HE-LHC will cover the complete parameter region on the plane.
The underlying figure is taken from Ref.~\cite{Blanke:2022kpi}.
} 
\end{center}
\end{figure}
The black contour in Fig.\,\ref{Fig:mutau2HDM_summary} shows the value of $\rho_e^{\mu\tau}\rho_e^{\tau\mu}$ required to explain the central value of the muon $g-2$ anomaly with the assumption of $|\rho_e^{\mu\tau}|=|\rho_e^{\tau\mu}|$.
For given masses $m_{H,A}$, the NP effect in $a_\mu$ is large if both couplings $\rho_e^{\mu\tau}$ and $\rho_e^{\tau\mu}$ are large while the Landau pole resides at a high-energy scale.
As we are interested in the heaviest scenario, we thus set $|\rho_e^{\mu\tau}|=|\rho_e^{\tau\mu}|$.
The $\tau$ decay constraint, which mainly comes from the charged-scalar tree-level correction, is depicted by the blue region.
It is known that the systematic uncertainty is already the dominant one \cite{Belle-II:2018jsg}.
The Belle II experiment will improve the sensitivity of the $\tau$ decays in the future while its detailed prospect is not available.
It should be added that the constraint from $Z$-boson decays is weaker in this scenario \cite{Omura:2015nja}.

The dashed magenta contours in Fig.\,\ref{Fig:mutau2HDM_summary} show the HL-LHC reach, where the region to the left of the contour can be probed.
A future high energy (HE)-LHC taking data at $\sqrt{s}=27$\,TeV with 3\,ab$^{-1}$ can cover the entire region in the plane depicted in Fig.\,\ref{Fig:mutau2HDM_summary}, see Ref.~\cite{Blanke:2022kpi} for more details.
Since the main production mechanism is via EW processes which are insensitive to $\rho_e^{\mu\tau}$ and $\rho_e^{\tau\mu}$, the result does not change drastically even if the anomaly needs to be explained at the $-1\,\sigma$ level.

We note that decays into gauge bosons such as $H^+\to W^+ H$ are not kinematically allowed in the $\mathcal{O}$(1)\,TeV scenario since the difference of the squared masses is $\mathcal{O}(v^2)$, see~Eq.~(\ref{Eq:mass_l5}).
Even if $\Delta a_\mu$ decreases, the distinctive signal cross section is controlled by the gauge coupling and thus the proposed $\mu^\pm\mu^\pm\tau^\mp\tau^\mp$ final state would serve as a smoking gun signal.

\section{Summary and discussion}
\label{sec:summary}
The current deviation in the muon anomalous magnetic moment could be a long-awaited hint of new physics.
In this review article, we revisited the muon $g-2$ anomaly within two-Higgs-doublet models.
Despite the fact that the 2HDM is one of the simplest extensions of the SM, the model has very rich phenomenology and constitutes the scalar sector of several UV-completing models.
In addition to the $\mathbb{Z}_2$-based type-X 2HDM, a flavor-aligned 2HDM and the $\mathbb{Z}_4$-based muon-specific 2HDM and $\mu\tau$-flavor violating 2HDM were considered.
These models have been known to be a possible solution to the muon $g-2$ anomaly.
We updated the collider constraints which give crucial bounds and clarified the available parameter space.

We found that due to the updated constraint from $B_s \to \mu^+ \mu^-$ the up-type Yukawa coupling in the FA2HDM cannot be large unless the contribution is cancelled by a down-type Yukawa coupling.
If we do not rely on this tuning, the Barr-Zee contribution with a top quark cannot explain more than $5\%$ of the discrepancy.
Therefore, the scenario is effectively the same as the type-X 2HDM in this case.

Although tau-rich signatures at the LHC provide a distinctive test of the type-X 2HDM and also the FA2HDM interpretation, a detailed Run~2 analysis has not yet been performed for this signature.
Based on the latest chargino-neutralino searches with the Run~2 full data, we found that the muon $g-2$ anomaly favored parameter region is certainly excluded in the type-X 2HDM and FA2HDM.
We emphasize that even if we employ the cancellation in $B_s\to \mu^+ \mu^-$, as long as tauonic and bosonic scalar decays are dominant, the interesting region is excluded.
Therefore the conventional $\mathbb{Z}_2$-based scenario is no longer viable for the explanation of the muon $g-2$ anomaly.

We also revisited the $\mu$2HDM and found that the Run~2 data pushed up the lower mass bound by 200\,GeV compared to the previous analysis and the model encounters the Landau pole at less than 5\,TeV if the central value of the muon $g-2$ anomaly is required.
We also found that the complete $1\,\sigma$ region satisfying $\Lambda_{\rm{LP}}\ge 5\,$TeV can be probed with a near future $500\,\rm{fb}^{-1}$ data set, which approximately corresponds to the integrated luminosity of the Run\,3 full data.
The status and prospect of the $\mu\tau$2HDM was also summarized.
In this model a distinctive $\mu^\pm\mu^\pm\tau^\mp\tau^\mp$ final state at the LHC is a key prediction.
We found that the HL-LHC can probe scenarios with scalars of up to 1.1\,TeV.
Together with the Landau pole constraint, a future upgrade of the HL-LHC to energies of $27\,$TeV could cover the complete parameter space relevant for the muon $g-2$ anomaly.
The summary of the relevant flavor and collider constraints for those four kinds of 2HDMs is shown in Table \ref{tab:ModelNRC}.

In this review, we focused on a simple extension that features only one additional Higgs doublet.
It is known that a further extension, \eg, 2HDM$+$vector-like lepton, can explain the muon $g-2 $ anomaly (see Ref.~\cite{Chun:2020uzw} for instance), since the heavier vector-like lepton mass can be used to flip the chirality, leading to a much larger mass enhancement factor.\footnote{However, it is known that with only vector-like leptons one can explain the muon $g-2$ anomaly even without introducing an additional scalar doublet.}
Further collider searches for the vector-like lepton scenario will shed light on a possible realization of these extended models \cite{CMS:2019lwf,ATLAS:2022yhd,CMS:2022nty}.

\section*{Acknowledgements}
S.\,I. appreciates the warm hospitality of Yuji Omura and Kindai University and where he started this project.
The research of S.\,I. and M.\,L. is supported by the Deutsche Forschungsgemeinschaft (DFG, German Research Foundation) under grant 396021762-TRR\,257.
T.\,K. is supported by the Grant-in-Aid for Scientific Research from the Ministry of Education, Culture, Sports, Science, and Technology (MEXT), Japan, No.\,21K03572.
The work of T.\,K. is supported by the Japan Society for the Promotion of Science (JSPS) Core-to-Core Program, No.\,JPJSCCA20200002.
M.T. is supported by the Fundamental Research Funds for the Central Universities, the One Hundred Talent Program of Sun Yat-sen University, China, and by the JSPS KAKENHI Grant, the Grant-in-Aid for Scientific Research\,C, Grant No.\,18K03611.


\appendix
\section{Explicit formulae}

In this appendix,
we collect all relevant formulae required in the analyses of this article.

\subsection{Muon \texorpdfstring{$g-2$}{g-2}}
\label{app:muong-2}

The leading-order Feynman diagrams contributing to possible solutions of the muon $g-2$ anomaly are shown in Fig.\,\ref{Fig:diag-2}.

The one-loop flavor conserving contribution is given as \cite{Abe:2017jqo}
\begin{align}
    \delta a_\mu^{\mu}\simeq \frac{G_F v^2 m_\mu^2}{8\sqrt{2}\pi^2}(\rho_{e}^{\mu\mu})^2 \left[ \frac{1}{m_A^2}\left(\frac{11}{6}+\log\frac{m_\mu^2}{m_A^2} \right)-\frac{1}{m_H^2}\left(\frac{7}{6}+\log\frac{m_\mu^2}{m_H^2} \right)-\frac{1}{6m_{H^\pm}^2}\right]\,.
\end{align}
The contributions from $A$ and $H$ have opposite sign and $H^\pm$ always gives a negative yet tiny contribution to the muon $g-2$.
A positive shift is realized for $m_H \lesssim m_A$.
 
Next, the one-loop $\mu\tau$-flavor violating contribution receives the tau-mass chirality enhancement factor as \cite{Omura:2015nja},
\begin{align}
\delta a_\mu^{\tau}
\simeq\frac{m_\mu^2 }{16\pi^2}\rho_{e}^{\mu\tau}\rho_{e}^{\tau\mu}\frac{m_\tau}{m_\mu}\left( \frac{\log{\frac{m_H^2}{m_\tau^2}}-\frac{3}{2}}{m_H^2}-\frac{\log{\frac{m_A^2}{m_\tau^2}}-\frac{3}{2}}{m_A^2}\right)\,,
\label{eq:gm2}
\end{align}
where the $H^\pm$-loop contribution does not have the tau-mass enhancement and can thus be neglected.
Again the contributions from $A$ and $H$ have opposite signs.
A positive contribution to the muon $g-2$
requires $m_H\le m_A$ for $\rho_{e}^{\mu\tau}\rho_{e}^{\tau\mu}> 0 $ and $m_H\ge m_A$ for $\rho_{e}^{\mu\tau}\rho_{e}^{\tau\mu}< 0 $.

Finally, the contribution from the two-loop Barr-Zee diagram is given as \cite{Ilisie:2015tra}
\begin{align}
\delta a_\mu^{\rm BZ}\simeq \frac{\alpha m_\mu}{16 \pi^3} \left\{\frac{4\rho_u^{tt}\rho_e^{\mu\mu}}{3m_t}\left[F_1(x_{tH}) - F_2(x_{tA})\right]
+\frac{\rho_e^{\tau\tau}\rho_e^{\mu\mu}}{m_\tau}\left[F_1(x_{\tau H}) + F_2(x_{\tau A})\right]\right\}\,,
\end{align}
where $x_{f\phi}=m_f^2/m_\phi^2$ and the loop functions are defined as 
\begin{align}
F_1(x)=&x\int^1_0 dy \frac{2y(1-y)-1}{x-y(1-y)} \log \left[ \frac{x}{y(1-y)} \right]\,,\\
F_2(x)=&x\int^1_0 dy \frac{1}{x-y(1-y)} \log \left[ \frac{x}{y(1-y)} \right]\,.
\end{align}

\subsection{Tau decays}
\label{app:tau_decay}

The treatment of the constraint arising from tau-lepton decays is crucial in order to judge the type-X 2HDM interpretation.
There are five precision observables, the correlations among which should be taken into account.
The HFLAV constraints on the tau effective couplings are summarized as \cite{HFLAV:2022pwe},
\begin{align}
   \biggl{(} \frac{g_\tau}{g_\mu}\biggl{)}_{\tau}&=1.0009\pm0.0014\,,~~
   \biggl{(} \frac{g_\tau}{g_e}\biggl{)}_{\tau}=1.00027\pm0.0014\,,~~\nonumber\\
   \biggl{(} \frac{g_\mu}{g_e}\biggl{)}_{\tau}&=1.0019\pm0.0014\,,~~
   \biggl{(} \frac{g_\tau}{g_\mu}\biggl{)}_{\pi}=0.9959\pm0.0038\,,\nonumber\\
   \biggl{(} \frac{g_\tau}{g_\mu}\biggl{)}_{K}&=0.9855\pm0.0075\,,
\end{align}
where the symmetric correlation matrix is given by
\begin{align}
\begin{pmatrix}
 1 &  &  &  & \\
 0.51 & 1 &  &  & \\
 -0.50 & 0.49 & 1 &  & \\
 0.16 & 0.18 & 0.01 & 1 & \\
 0.12 & 0.11 & -0.01 & 0.07 & 1\\
\end{pmatrix}\,.
\end{align}
The effective couplings for leptonic tau decays are defined as
\begin{align}
\left(\frac{g_\tau}{g_\mu}\right)^2_\tau \propto \frac{\Gamma(\tau \to e \nu_\tau\overline\nu_e)}{\Gamma(\mu \to e \nu_\mu\overline\nu_e)}\,, \quad 
        \left(\frac{g_\tau}{g_e}\right)^2_\tau \propto \frac{\Gamma(\tau \to \mu \nu_\tau\overline\nu_\mu)}{\Gamma(\mu \to e \nu_\mu\overline\nu_e)}\,,\quad 
    \left(\frac{g_\mu}{g_e}\right)^2_\tau \propto \frac{\Gamma(\tau \to \mu \nu_\tau\overline\nu_\mu)}{\Gamma(\tau \to e \nu_\tau\overline\nu_e)}\,,
\end{align}
while the ones for hadronic tau decays ($h=\pi, K$) are defined as
\begin{align}
    \left( \frac{g_{\tau}}{g_{\mu}}\right)^2_h \propto \frac{\Gamma(\tau \to h \nu_\tau)}{\Gamma(h \to \mu \overline{\nu}_\mu)}\,.
\end{align}
These ratios are normalized by the phase spaces so that the SM predictions are 1.

Tree-level and one-loop corrections to the tau effective couplings are calculated in Refs.~\cite{Krawczyk:2004na,Abe:2015oca}.
The contributions are given by 
\begin{align}
   \left( \frac{g_\tau}{g_\mu}\right)_{\tau,\,\pi,\,K}&\simeq 1+\delta_\tau^{\rm{loop}} -\delta_\mu^{\rm{loop}},\\
   \left( \frac{g_\tau}{g_e}\right)_{\tau}&\simeq 1+\delta_\tau^{\rm{tree}}+\delta_\tau^{\rm{loop}} -\delta_\mu^{\rm{loop}},\\
   \left( \frac{g_\mu}{g_e}\right)_{\tau}&\simeq 1+\delta_\tau^{\rm{tree}},
\end{align}
with
\begin{align}
       \delta_\tau^{\rm{tree}}&=\frac{1}{2}\left[-\frac{v^2} {m_{H^\pm}^2}\rho_e^{\mu\mu}\rho_e^{\tau\tau} \frac{m_\mu g(m_\mu^2/m_\tau^2)}{m_\tau f(m_\mu^2/m_\tau^2)}+\frac{v^4}{16 m_{H^\pm}^4}\left(\rho_e^{\mu\mu}\rho_e^{\tau\tau}\right)^2\right]\,,\\
    \delta_\ell^{\rm{loop}}& =\frac{(\rho_e^{\ell\ell})^2}{32\pi^2}\left\{1+\frac{1}{4}\left[h\left(\frac{m_{A}^2}{m_{H^\pm}^2}\right)+h\left(\frac{m_{H}^2}{m_{H^\pm}^2}\right)\right] \right\} \,.
\end{align}
A contribution from the tree-level $H^\pm$ exchange in $\tau \to \mu \nu_{\tau}\overline{\nu}_{\mu}$ is represented by $\delta_{\tau}^{\rm tree}$, while radiative corrections to the $W$-$\ell$-$\nu_\ell$ couplings are denoted by $\delta_{\ell}^{\rm loop}$.
The loop functions are given by 
\begin{align}
f(x)&=1-8x+8x^3-x^4-12x^2\log{x}\,,\\
g(x)& =1+9x-9x^2-x^3+6x(1+x)\log{x}\,,\\ h(x)&=\frac{1+x}{1-x}\log{x}\,.
\end{align}  

To investigate the exclusion region from tau decays, the $\chi^2$ is constructed based on these five observables including their correlations.
For the $\mu\tau$2HDM, the tau decay constraints are less significant due to the heavy additional scalar masses, even if one includes the Michel parameters \cite{Tobe:2016qhz}.
\subsection{\texorpdfstring{$Z$}{Z} decays}
\label{app:Z_decay}
Sizable lepton Yukawa couplings change the $Z$-boson vertices at the one-loop level \cite{Abe:2015oca,Abe:2019bkf,Chun:2016hzs}.
The corrections to the ratios of the leptonic decay widths are given by
\begin{align}
\frac{\Gamma(Z\to \tau^+\tau^-)}{\Gamma(Z\to e^+ e^-)}&\simeq 1+2\frac{g_L^e{\rm{Re}}(\delta g_L)+g_R^e{\rm{Re}}(\delta g_R)}{(g_L^{e})^2+(g_R^{e})^2}\,,\\
\frac{\Gamma(Z\to \mu^+\mu^-)}{\Gamma(Z\to e^+ e^-)}&\simeq 1+2\frac{g_L^e{\rm{Re}}(\delta g_L)+g_R^e{\rm{Re}}(\delta g_R)}{(g_L^{e})^2+(g_R^{e})^2}\left(\frac{\rho_e^{\mu\mu}}{\rho_e^{\tau\tau}}\right)^2\,,
\end{align}
with the vertex corrections 
\begin{align}
\delta g_{L}&=\frac{(\rho_e^{\tau\tau})^2}{32 \pi^2}\biggl{\{} -\frac{1}{2} \left[B_Z(x_A)+B_Z(x_H)+4C_Z(x_A, x_H)\right] \nonumber\\
&\quad +s_W^2\left[B_Z(x_A)+B_Z(x_H)+\tilde{C}_Z(x_A)+\tilde{C}_Z(x_H)   \right]\biggl{\}}\,,\\
\delta g_{R}&=\frac{(\rho_e^{\tau\tau})^2}{32 \pi^2}\biggl{\{} \frac{1}{2} \left[4C_Z(x_A,x_H)-4C_Z(x_{H^\pm},x_{H^\pm})+
2\tilde{C}_Z(x_{H^\pm})-\tilde{C}_Z(x_{A})-\tilde{C}_Z(x_{H})\right]\nonumber\\
&\quad +s_W^2\left[ B_Z(x_A)+B_Z(x_H)+2B_Z(x_{H^\pm})+4C_Z(x_{H^\pm},x_{H^\pm})+\tilde{C}_Z(x_A)+\tilde{C}_Z(x_H) \right]\biggl{\}}\,,
\end{align} 
where $x_\phi = m_\phi^2/m_Z^2$, $s_W \equiv \sin \theta_W$, and $g_{L,R} = T_3 - Q s_W^2$ ($g^e_L\simeq -0.27$ and $g_R^e\simeq 0.23$).
The loop functions are defined as 
\begin{align}
    B_Z(x)&=-\frac{1}{2\bar\epsilon}-\frac{1}{4}+\frac{1}{2}\log x\,,\\
    C_Z(x,y)&=\frac{1}{4\bar\epsilon}-\frac{1}{2}\int^1_0dz_1 \int^{z_1}_0 dz_2 \,\log\left[ z_1z_2+y(1-z_1)+(x-1)z_2\right]\,,\\
    \tilde{C}_Z(x)&=\frac{1}{2\bar\epsilon}+\frac{1}{2}-x\left(1+\log x\right)+x^2\left[ \log x \log(1+x^{-1})-\rm{Li}_2(-x^{-1})\right]\nonumber\\
    &\quad -\frac{{\rm{i}}\pi}{2}\left[ 1-2x+2x^2 \log(1+x^{-1})\right]\,,
\end{align}
where $\rm{Li}_2$ denotes the di-logarithm function and the 1/$\bar\epsilon$ poles cancel in the sum.
We confirm that the forward-backward asymmetry and the tau polarization asymmetry in $Z\to\tau^+\tau^-$ give less stringent constraints.

For $\mu\tau$2HDM, $(\rho_e^{\tau\tau})^2$ is replaced by $(\rho^{\mu \tau}_e)^2 $ in $\delta g_L$ and by $(\rho^{\tau \mu}_e)^2 $ in $\delta g_R$, respectively~\cite{Abe:2019bkf}.
For $|\rho^{\mu\tau}_e|=|\rho^{\tau\mu}_e|$ the constraint is less stringent.

\subsection{\texorpdfstring{$B_s\to\mu^+ \mu^-$}{Bs to mumu}}
\label{app:Bsmumu}

\begin{figure}[t]
    \begin{center}
\includegraphics[width=14em]{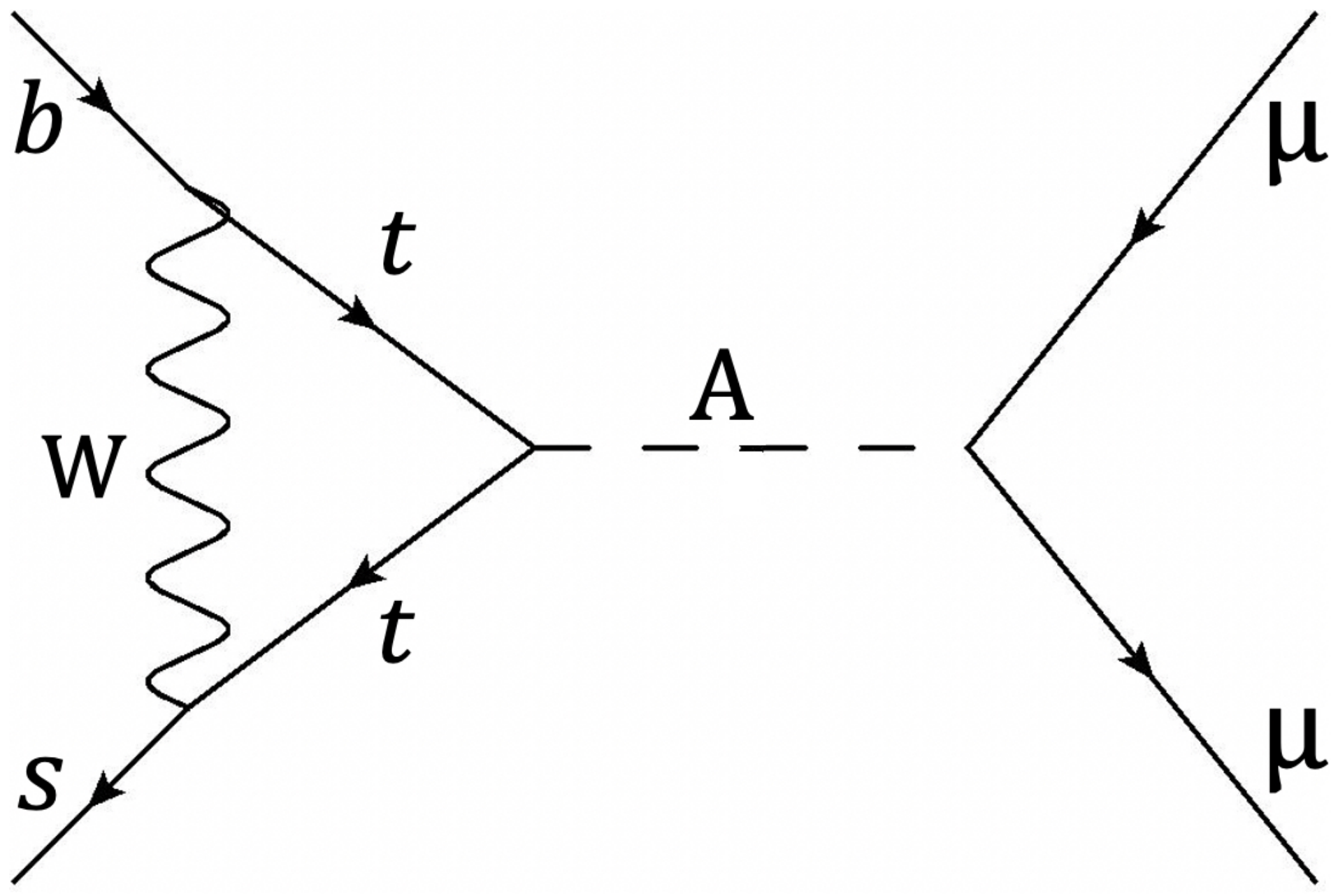}
\caption{
\label{Fig:dia_bsmumu}
The dominant Feynman diagram for $B_s\to \mu^+ \mu^-$
in the type-X 2HDM and FA2HDM.
} 
\end{center}
\end{figure}

In this section, we discuss the constraint from $B_s\to\mu^+ \mu^-$.
In Ref.~\cite{Li:2014fea}, the calculation of the full one-loop Wilson coefficients contributing to $B_s \to \mu^+ \mu^-$ has been performed within the FA2HDM.
The recent CMS result \cite{CMS:2022dbz} is consistent with the SM prediction
\cite{Bobeth:2013uxa,Beneke:2017vpq,Buras:2022wpw}.
Since the type-X 2HDM increases the branching ratio of $B_s \to \mu^+ \mu^-$, the recent shift of the experimental world average \cite{HFLAV2023January} weakens the $m_A$ bound stemming from $B_s \to \mu^+ \mu^-$ compared to the previous world average~\cite{CMS-PAS-BPH-20-003,Altmannshofer:2021qrr}.
This bound is relevant for the type-X 2HDM and FA2HDM,
since the dominant contribution comes from the one-loop diagram with the light CP-odd scalar mediation, shown in Fig.\,\ref{Fig:dia_bsmumu}.
We adopted the formulae from Ref.~\cite{Li:2014fea} and derived the constraint.

Since $\rho_e^{\tau\tau}$ is larger than $\rho_e^{\mu\mu}$, one might think that $B_s\to \tau^+ \tau^-$ could be a good decay process in order to probe the 2HDMs.
However, this does not hold and the NP sensitivity would be the same as $B_s\to \mu^+ \mu^-$, because the SM amplitude ($W$-box and $Z$-penguin diagrams) is also proportional to $m_\tau$.
Moreover, $B_s\to \tau^+ \tau^-$ has not yet been observed.

\subsection{Perturbative unitarity and vacuum stability}
\label{app:PUnVS}
In this section we discuss theoretical constraints imposed on the couplings in the scalar potential by the requirement of perturbative unitarity and the vacuum stability.
The constraints from perturbativity can be derived from the consideration of scattering amplitudes of the Higgs bosons.
Following Ref.~\cite{Lee:1977eg} where longitudinally polarized gauge bosons are replaced with the corresponding Nambu-Goldstone bosons, we only consider the scattering processes involving scalars and gauge bosons.
The full set of scattering amplitudes is expressed as a $22\times22$ matrix, which falls apart into four decoupled sub-matrices~\cite{Kanemura:1993hm,Akeroyd:2000wc,Arhrib:2000is,Ginzburg:2003fe,Ginzburg:2005dt,Horejsi:2005da,Abe:2015oca}.
The perturbative unitarity bound is imposed on the $12$ distinct eigenvalues of the matrix as
\begin{equation}
|e_j| < 8\pi \quad (j=1,\dots,12)\,,
\end{equation}
where 
\begin{align}
e_{1,2} &= \lambda_3 \pm \lambda_4\,, \ e_{3,4} = \lambda_3 \pm \lambda_5 \,, \ e_{5,6}=\lambda_3+2\lambda_4\pm3\lambda_5\,, \\
e_{7,8} &= \frac{1}{2}\left[ (\lambda_1+\lambda_2)\pm\sqrt{(\lambda_1-\lambda_2)^2+4\lambda_4^2}\right]\,, \\
e_{9,10} & = \frac{1}{2}\left[3(\lambda_1+\lambda_2)\pm\sqrt{9(\lambda_1-\lambda_2)^2+4(2\lambda_3+\lambda_4)^2}\right] \,,\\
e_{11,12} & = \frac{1}{2}\left[(\lambda_1+\lambda_2)\pm\sqrt{(\lambda_1-\lambda_2)^2+4\lambda_5^2}\right] \,,
\end{align}
where all $\lambda_j$ are running couplings.
Here, the contributions from $\lambda_6$ and $\lambda_7$ are discarded; $\lambda_6=0$ is fixed by the Higgs alignment condition and $\lambda_7$ is suppressed by large $\tan\beta$, see \eq{eq:l7}.
We define our cutoff scale $\Lambda_{\rm{LP}}$ as the minimum scale at which either the vacuum stability or perturbative unitarity condition breaks down when evolving the couplings with the RGEs from an input scale to a high-energy scale.
It should be noted that the quartic couplings are also bounded from below by the conditions \cite{Maniatis:2006fs,Ferreira:2009jb},
\begin{equation}
\lambda_1, \lambda_2 \geq 0\,, \quad 
\sqrt{\lambda_1\lambda_2}+\lambda_3 \geq 0\,,\quad
\sqrt{\lambda_1\lambda_2}+\lambda_3+\lambda_4-|\lambda_5| \geq0\,,
\end{equation}
but these conditions are always satisfied in the parameter region of our interest.

\subsection{Renormalization group equations}
\label{app:beta}

The RGEs of the scalar quartic couplings in the $\mu$2HDM and $\mu\tau$2HDM are given in the form of
\begin{equation}
\frac{d\lambda_j}{d\log\mu} = \frac{\beta_{\lambda_j}}{(4\pi)^2}\,,
\end{equation}
where $\mu$ is the renormalization scale.
The RGE running effect is important for these two 2HDMs since the scalars are heavy and thus the Yukawa couplings are large.

At the one-loop level, the beta functions $\beta_{\lambda_j}$ of the scalar potential in \eq{eq:Higgsbasis} are given by~\cite{Ferreira:2009jb,Goudelis:2013uca}
\begin{align}
\beta_{\lambda_1} 
   &=12\lambda_1^2+4\lambda_3^2+4\lambda_3\lambda_4+2\lambda_4^2+2\lambda_5^2 + 24 \lambda_6^2
       + \frac{3}{4}(3g^4+g^{\prime4}+2g^2g^{\prime2}) 
       \nonumber\\
   &\quad - 3\lambda_1(3g^2+g^{\prime2}) +12\lambda_1y_t^2 - 12y_t^4\,, \\
\beta_{\lambda_2} 
   &=12\lambda_2^2+4\lambda_3^2+4\lambda_3\lambda_4+2\lambda_4^2+2\lambda_5^2+24 \lambda_7^2 
       + \frac{3}{4} (3g^4+g^{\prime4}+2g^2g^{\prime2}) - 3\lambda_2(3g^2+g^{\prime2})
       \nonumber\\
   &\quad  +4\lambda_2[(\rho_e^{\ell\ell})^2+(\rho_e^{\mu\tau})^2+(\rho_e^{\tau\mu})^2]
      - 4[(\rho_e^{\ell\ell})^4+(\rho_e^{\mu\tau})^4+(\rho_e^{\tau\mu})^4]\,, \\
\beta_{\lambda_3} 
   &=2(\lambda_1+\lambda_2)(3\lambda_3+\lambda_4) 
       + 4\lambda_3^2+2\lambda_4^2+2 \lambda_5^2 +4 \lambda^2_6 +16 \lambda_6 \lambda_7 + 4 \lambda_7^2 - 3\lambda_3(3g^2+g^{\prime2}) 
      \nonumber\\
   &\quad  +\frac{3}{4}(3g^4+g^{\prime4}-2g^2g^{\prime2})
       + 2\lambda_3[3y_t^2+(\rho_e^{\ell\ell})^2+(\rho_e^{\tau\mu})^2+(\rho_e^{\mu\tau})^2]\,, \\
\beta_{\lambda_4} 
   &=2\lambda_4(\lambda_1+\lambda_2+4\lambda_3+2\lambda_4)+8\lambda_5^2 + 10 \lambda_6^2 + 4 \lambda_6 \lambda_7 + 10 \lambda_7^2
       +3g^2g^{\prime2} \nonumber\\
   &\quad -3\lambda_4(3g^2+g^{\prime2})+ 2\lambda_4[3y_t^2+(\rho_e^{\ell\ell})^2+(\rho_e^{\tau\mu})^2+(\rho_e^{\mu\tau})^2]\,, \\
\beta_{\lambda_5} 
   &=2\lambda_5(\lambda_1+\lambda_2+4\lambda_3+6\lambda_4) +10 \lambda_6^2 + 4 \lambda_6 \lambda_7 + 10 \lambda_7^2
       - 3\lambda_5(3g^2+g^{\prime2})\nonumber\\
   &\quad + 2\lambda_5[3y_t^2+(\rho_e^{\ell\ell})^2+(\rho_e^{\tau\mu})^2+(\rho_e^{\mu\tau})^2]\,,\\
\beta_{\lambda_6} 
   &= 12 \lambda_1 \lambda_6 + 6 \lambda_3 (\lambda_6 + \lambda_7)+8 \lambda_4 \lambda_6 + 4 \lambda_4 \lambda_7 + 10\lambda_5 \lambda_6 + 2 \lambda_5 \lambda_7
       - 3\lambda_6(3g^2+g^{\prime2})\nonumber\\
   &\quad + 3\lambda_6[3y_t^2+(\rho_e^{\ell\ell})^2+(\rho_e^{\tau\mu})^2+(\rho_e^{\mu\tau})^2]\,,\\
\beta_{\lambda_7} 
   &= 12 \lambda_2 \lambda_7 + 6 \lambda_3 (\lambda_6 + \lambda_7)+4 \lambda_4 \lambda_6 + 8 \lambda_4 \lambda_7 + 2\lambda_5 \lambda_6 + 10 \lambda_5 \lambda_7
       - 3\lambda_7(3g^2+g^{\prime2})\nonumber\\
   &\quad + \lambda_7 [3y_t^2+(\rho_e^{\ell\ell})^2+(\rho_e^{\tau\mu})^2+(\rho_e^{\mu\tau})^2]\,,
\end{align}
where the Yukawa couplings are defined by $y_f = \sqrt{2} m_f/v$.
The RGEs of the gauge and Yukawa couplings are defined in the same way, with the beta functions given by 
\begin{align}
\beta_{g_j} & = b_j g_j^3\,, \quad b_j=\{7,-3,-7\}  \quad (g_j=\{g',g,g_s\})\,,\\
\beta_{y_t} & = y_t \left( -\frac{17}{12} g^{\prime2} - \frac{9}{4} g^2 - 8g_s^2  +\frac{9}{2} y_t^2 \right)\,,\\
\beta_{\rho_e^{\ell\ell}} & = \rho_e^{\ell\ell} \left[ \frac{5}{2}(\rho_e^{\ell\ell})^2-\frac{9}{4} \left( \frac{5}{3}g^{\prime\,2}+g^2\right) \right]\,,\\
\beta_{\rho_e^{\tau\mu}} & = \rho_e^{\tau\mu} \left[ (\rho_e^{\mu\tau})^2+\frac{5}{2}(\rho_e^{\tau\mu})^2-\frac{9}{4} \left( \frac{5}{3}g^{\prime\,2}+g^2\right) \right]\,,\\
\beta_{\rho_e^{\mu\tau}} & = \rho_e^{\mu\tau} \left[ (\rho_e^{\tau\mu})^2+\frac{5}{2}(\rho_e^{\mu\tau})^2-\frac{9}{4} \left( \frac{5}{3}g^{\prime\,2}+g^2\right) \right]\,.
\end{align}

\bibliographystyle{utphys28mod}

\bibliography{ref}

\end{document}